\documentclass[life,article,moreauthors,pdftex,a4paper]{mdpi}

\lastpage{x}
\doinum{10.3390/------}
\pubvolume{xx}
\pubyear{2013}
\history{Received: xx / Accepted: xx / Published: xx}

\usepackage{amssymb}

\Title{Disk evolution, element abundances and cloud properties of
 young  gas giant planets}

\Author{Ch.~Helling $^{1,\!\displaystyle *}$, 
        P.~Woitke $^{1}$,
        P.~Rimmer $^{1}$,
        I.~Kamp $^{2}$,
        W.-F.~Thi $^{3}$,
        R.~Meijerink $^{2,4}$}

\address{
$^{1}$ SUPA, University of St Andrews, School of Physics \& Astronomy, 
       North Haugh, KY16 9SS, St Andrews, UK\\
$^{2}$ Kapteyn Astronomical Institute, Postbus 800, 9747 AV Groningen, 
       The Netherlands\\
$^{3}$ Laboratoire d'Astrophisique de Grenoble, CNRS/Universit{\'e} 
       Joseph Fourier (UMR5571) BP 53, F-38041 Grenoble cedex 9, France\\
$^{4}$   Leiden Observatory, Leiden University, P. O. Box 9513, 2300 RA Leiden, The
Netherlands   
       }

\corres{ch80@st-and.ac.uk}

\abstract{
\nolinenumbers 
 We discuss the chemical pre-conditions for planet formation, in terms
 of gas and ice abundances  in a protoplanetary disk, as function
 of time and position, and the resulting chemical composition and
   cloud properties in the atmosphere when young gas giant planets
   form, in particular discussing the effects of unusual, non-solar
   carbon and oxygen abundances.
  Large deviations between the abundances of the host star and its
  gas giants seem likely to occur if the planet formation follows the
  core-accretion scenario.  These deviations stem from  the
  separate evolution of gas and dust in the disk, where the dust forms
  the planet cores, followed by the final run-away accretion of the
left-over gas. This gas  will contain only traces of elements
 like C, N and O, because those elements have frozen out as ices.
 {\sc ProDiMo} protoplanetary disk models are used to predict the
 chemical evolution of gas and ice in the midplane. We find that
 cosmic rays play a crucial role in slowly un-blocking the CO, where
 the liberated oxygen forms water, which then freezes out quickly.
 Therefore, the C/O ratio in the gas phase is found to gradually
 increase with time, in a region bracketed by the water and CO
 ice-lines.   In this regions, C/O is found to approach unity
 after about 5\,Myrs, scaling with the cosmic ray ionisation rate
 assumed.
 We then explore how the atmospheric chemistry and cloud
   properties in young gas giants are affected when the non-solar C/O
   ratios predicted by the disk models are assumed. The {\sc
   Drift} cloud formation model is applied to study the formation of
 atmospheric clouds under the influence of varying premordial element
 abundances and its feedback onto the local gas.
We demonstrate that element depletion by cloud formation plays a
crucial role in converting an oxygen-rich atmosphere gas into
carbon-rich gas when non-solar, premordial element abundances are
considered as suggested by disk models. }

\begin{document}
\nolinenumbers  


\section{Introduction}

\noindent Element abundances are critical parameters to predict the
atmospheric composition of exoplanets and to understand their
formation and evolution, including potentially the emergence of life.
Extrasolar gas giants are commonly assumed to have elemental
abundances similar to those of their host stars. These stars
themselves can be reasonably well measured,
for example (in case of the Sun) by high-resolution spectroscopy in
combination with time-dependent numerical simulations of the
photosphere, meteorite studies, or astroseismology \cite{asp09}.
However, when considering the process of planet formation in a
protoplanetary disk, which involves a segregation of gas, dust and ice
phases, the assumption that the element mix of the host star must be
the same as in the gas giants' atmospheres becomes questionable. 
  This has far-reaching consequences for the spectroscopic analysis of
  planetary spectra, including the search for bio-signatures
\cite{hwt08,wit09,bi2013}. Following the standard core-accretion model
of planet formation \cite{Wetherill96,Ida2004}, the refractory
elements are initially present mostly in form of $\mu$m-sized dust
particles, which undergo a complex evolution 
eventually leading to km-sized planetesimals. The planetesimals are
gravitationally attracted to each other, and collide to form larger
parental bodies that later become planetary cores \cite{Kokubo1998,Fortier2013}.

At the end of the  evolution from dust to planet cores, the
planet feeding zone is expected to be mostly devoid of smaller dust
particles \cite{Dullemond2005}, and the remaining gas in the planet
feeding zone is expected to contain only minor traces of refractory
elements.  Elements which are able to form ices on the surface of the
refractory grains  in protoplanetary disks, in particular oxygen,
carbon and nitrogen, will also be depleted to an extent,  depending
  on local temperature, though less than the refractory elements.
The ices play an essential role in the dust growth process as ``glue''
or ``cement''  during planet formation \cite{Bridges1996}.  The
dust particles have been in contact with the gas in the disk for
$>\!10^6$\,yrs, which is certainly long enough to cause most of the
gaseous oxygen in the disk midplane to form H$_2$O ice outside the
water ``ice-line'' and for most of the gaseous carbon to form CO ice
outside the CO ``ice-line'' \cite{Chaparro2012a}. The
elements bound in those ices should then rather follow the dust than
the gas dynamical evolution. Only at the very end of the planet
formation process, the overwhelming majority of the mass of the gas
giant will be accreted onto the proto-planet in a rapid run-away phase
\cite{Pollack1996}, using up the remaining gas in the planet feeding
zone and  possibly forming a gap. The
timescale for gas accretion onto the proto-planet is about two orders of magnitude
shorter than the growth timescale of the solid core \cite{Machida2010}. At this late
stage, the gas should contain only traces of refractory elements, and
possibly also very little amounts of the ``ice elements'' O, C and N,
depending on local temperature, i.e.\ position in the disk.   The
  resulting planetary atmosphere will hence be extremely metal-poor in
  the first place. Late bombardment with left-over
  planetesimals \cite{Zhou2007,Bergin2013} will cause an element
  re-enrichment leading once more to a change of the atmospheric
  composition. The opacity of the atmosphere surrounding the
  planetary core plays an important role for the critical mass that
  inhibits further accretion \cite{hase2014}.

Thus, the formation of gas giants via core-accretion means that,
first, the ice and volatile elements segregate. Then the gas and
icy dust evolves separately. Finally, the icy dust and the gas combine
in a specific order, forming a planet.
It would be a strange coincidence if all these complicated processes
would result in gas giant surface element abundances that resemble
those of their host stars. We should rather expect a large variety of
the atmospheric element abundances of gas giants, depending on when
and where the planets form. 


In this paper, we first study the segregation and evolution of
gas and ice  in protoplanetary disk models to predict the element
abundances of the gas that will finally be accreted onto the
proto-planets (Sect.~\ref{s:disk}). We show that the resulting
carbon-to oxygen ratio (C/O) is expected to be larger than the
primordial value, and increase further with time, in particular
between the water snowline ($\approx\!150$\,K) and the CO ice-line
($\approx\!20$\,K), where mostly water freezes out. Similar results
have been recently obtained in \cite{Oberg2011}, who schematically
discussed the relative segregation of carbon, oxygen, and nitrogen in
the disk, causing the C/O and C/H ratios to differ significantly from
those of the host stars. Gaseous C/O ratios close to unity between the
water and CO ice-lines, mainly driven by the formation of water, CO
and CO$_2$ ices \cite{Oberg2011}.  The time-dependent ice
  composition in the midplane of T\,Tauri disks has been studied by
\cite{Chaparro2012a,Chaparro2012b}, who found agreement with measured
chemical compositions of comets in their model, after long integration
times (10\,Myrs) at 10\,AU, using a relatively high cosmic ray ionization rate of
$5\times10^{-17}\rm\,s^{-1}$ and a sophisticated treatment of the
secondary cosmic ray induced photo reactions, the rates of which are
enhanced due to an increased ratio of UV gas absorption with respect
to dust absorption, driven by dust growth with respect to interstellar
conditions.

 We expect the metal-poor gas in protoplanetary disks to lead to
  unusual element abundances in planets, in particular the element
  abundances in gas giant atmospheres, although the dynamical details
  of the actual planet formation process need further investigation
  \cite{hase2014}.  Tentative detections of carbon-rich planets have
been announced, concerning the planets WASP\,12b and 55\,Cancri\,e,
respectively \cite{Madhusudhan2011, Madhusudhan2012}.
However, 65 orbits of WFC3-IR grism observations could not find any
evidence for $\rm C/O\!>\!1$ in the atmosphere of WASP\,12b
\cite{Swain2013}, as was reported \cite{Madhusudhan2011}. And in
the case of 55\,Cancri\,e, is was shown that the abundance
analysis of the host star 55\,Cancri \cite{Delgado2010} (used in
\cite{Madhusudhan2012}) was probably erroneous due to a unsuitable
choice of a zero-excitation oxygen line \cite{Nissen2013}.

Most of the presently known extrasolar planets orbit somewhat
metal-rich host stars \cite{buch12}. However, this simple relation
holds only for giant gas planets, but not for Neptune-sized
planets. It was argued that the Sun has a depletion of refractory
to volatile elements of about 20\% with respect to planet-free solar
twins \cite{mel09}. This abundance deficit roughly matches the mass of
the terrestrial planets.  One possible explanation for these
deficiencies is to assume an early formation of $\sim 10$\,km bodies,
which later formed the terrestrial planets before the majority of the
Sun's mass (excluding the $\sim 10$ km bodies) was accreted from the
proto-solar disk. The abundance peculiarities can also be found in
solar-like stars that are known to have close-in giant planets
\cite{mel09}.

 Our knowledge about the chemical composition of exoplanet atmospheres
 is prompted by transit observations of close-in planets,
 e.g.~\cite{cros2013, wil14, nik14}, or by bulk properties like global
 density estimates, e.g.~\cite{fort11}. Recent observation of the four
 directly imaged HD 8788 planets, however, suggest spectral diversity
 amongst co-eval objects of similar luminosity. The authors report on
 tentative detections of CH$_4$, C$_2$H$_2$, CO$_2$ and HCN
 \cite{opp13}.  A more complete understanding of exoplanet atmospheres
 hinges on the detailed atmosphere modeling that must include cloud
 formation, photochemistry and global circulation. The element
 abundances are essential parameters to all those models, and a simple
 scaling of a metallicity parameter, e.g. from \cite{mos13,cros2013},
 is far from realistic, as we will demonstrate in
 Sect.~\ref{s:cloudmod} of this paper. Planetary atmosphere chemistry
 does not only depend on the initial element abundances but also on
 the cloud formation process that depletes condensable elements and
 hence determines the remaining gas composition and radiative cooling
 processes. Cloud formation will impact all elements involved (Fe, Ti,
 Al, O, $\ldots$ \cite{hwt08, wit09}) and thereby change the C/O-ratio
 in such atmospheres (see Figs. 2 \& 3 in \cite{bi2013}).  Cloud
 layers have a large impact on the atmospheric structure and the
 spectral appearance of ultra-cool low-mass objects, like brown dwarfs
 and planets.  Thus, a direct abundance analysis of exoplanets like
 WASP\,12b, possibly with future instruments like JWST, must take
 these effects into account carefully.  Different cloud models make
 different predictions for the remaining gas-phase abundances
 resulting in different molecular abundances \cite{hell08b}. Clouds
 are expected to occur in a variety of physical phases and chemical
 compositions, depending on temperature and pressure, from cold icy
 hazes, over liquid droplets, to hot solid gemstones.  Beside
 temperature and pressure, the cloud formation process is controlled
 by the elemental composition of the atmospheric gas, which in turn is
 drastically reduced by the consumption of condensable elements into
 cloud particles and subsequent rain-out~\cite{hwt08}.


Section~\ref{s:disk} summarizes the disk chemistry model that we use to
  predict gas and ice elemental abundances in proto-planetary disks.
 Section~\ref{s:cloudmod} introduces our model for planetary
  atmospheres and cloud formation. Inspired by the results of the disk
  models, we consider unusual element abundances, in particular large
  ratios $\rm C/O\!\lesssim\!1$ in young gas giant atmospheres. In
Sect.~\ref{s:cloudeps}, we demonstrate that such non-standard oxygen
abundances have a strong impact on the atmospheric structure and cloud
properties in the atmosphere, like the cloud base, cloud particle
number densities, and mean grain sizes.  We further show that 
  condensation of oxygen-rich dust may cause the C/O ratio to tip over
  locally, $\rm C/O\!>\!1$, and we show that the abundances
  of astrobiologically interesting molecules like H$_2$O, CH$_4$,
  NH$_3$, C$_2$H$_2$, C$_2$H$_6$ may increase near the cloud top.

\section{Gas and ice abundances in protoplanetary disks}\label{s:disk}

\noindent In order to model the chemical composition of the gas and
ice in the disk as function of time, we follow a two-stage modeling
strategy. In {\sl disk modeling stage 1}, we simulate the chemistry in
the dark cores of molecular clouds by advancing our chemical rate
network under the corresponding temperature, density, and radiation
field conditions for an assumed lifetime of the dark core. The
resulting concentrations are then taken as initial conditions for the
disk simulations in {\sl disk modeling stage 2}.  This is a very much
simplified approach which ignores the complicated hydrodynamical star
and disk formation processes itself, as well as the changing
conditions in the disk during class-0 and class-I, although these
phases are relatively short.  Instead, we reset our clock at the
beginning of stage 2, and then advance our chemistry further under the
local temperature, density, dust, and radiation field conditions in
the disk.


More sophisticated models for the early disk phases, which evolve the
chemistry along streamlines from hydro-models for star and disk
formation, have been carried out by \cite{Visser2009}. In this paper,
we are interested in the long-term evolution of the chemistry, and
hence concentrate on class-II disks. Protoplanetary disks are
  observed to survive for about 3-10\,Myrs before they disperse
  \cite{hernandez08,Dent2013}. We evolve the disk chemistry for
  10\,Myrs, and even beyond, to study the asymptotic behavior towards
  kinetic chemical equilibrium which provides an interesting special
  case, providing additional understanding of the complete chemical
  paths for disk midplane evolution.

For our chemical simulations, we apply the radiation thermo-chemical
disk code {\sc ProDiMo} \cite{woi09}.  The 2D code consistently solves the
dust continuum radiative transfer, with the dust component in
radiative equilibrium, the gas heating and cooling balance (70 heating
processes, 64 cooling processes), the gas phase and ice chemistry, and
the non-LTE line transfer in class-II protoplanetary disks, with
recent updates described by \cite{woi11,Kamp2013}. The chemical model
consists of 12 elements (H, He, C, N, O, Mg, Si, S, Fe, Ne, Ar, PAH)
and 166 species, including vibrationally excited H$_2$, 30 ice species
with adsorption energies assumed as listed in Table~\ref{tab:Eads},
and 5 charging states of PAHs. Concerning the reaction rates, we have
experimented with two networks: the UMIST-2012 rates \cite{McElroy2013}
and the OSU-2010 rates with 2012 erratum
\cite{Harada2010,Harada2012}. In both cases, we select all reactions
among our selected species, and add the same set of additional
reactions for vibrationally excited H$_2$, UV ionization and
photo-dissociation (based on detailed UV cross sections from the Leiden
database \cite{vanDishoek2006}, coupled to the radiative transfer),
X-ray reactions \cite{Aresu2011}, PAH charging reactions, and ice
formation (adsorption, thermal desorption, UV desorption, cosmic ray
desorption), as well as H$_2$ formation on grains and some surface
chemistry, see \cite{Kamp2013}.

\begin{table}[!h]
\centering
\caption{Assumed adsorption energies for some important ices 
\cite[after][]{McElroy2013}.}
\label{tab:Eads}
\vspace*{-2mm}
\resizebox{\textwidth}{!}{
\begin{tabular}{|c|cccc|cccccc|ccc|}
\hline
ice species      & O & OH & H$_2$O & O$_2$ 
                 & CO & CO$_2$ & CH$_3$CO & C$_2$H$_2$ & CH$_3$ & CH$_4$
                 & N & N$_2$ & NH$_3$\\
\hline
$E_{\rm ads}$\,[K]& 800  & 2850 & 4800 & 1000 
                 & 1150 & 2990 & 4930 & 1400 & 1175 & 1090
                 &  800 & 790  & 5534 \\
\hline
\end{tabular}}
\end{table}

\subsection{Stage 1: the dense core simulations}

\noindent Stage 1 of our simulation is a simple one-point model for the
dense cores of molecular clouds, where we advance our chemical rate
network for a certain time, from initial atomic abundances typical
for the diffuse interstellar medium.

For the dense core conditions we adopt the following values as
recommended for TMC-1 by \cite{McElroy2013}: temperatures $T_{\rm
  gas}\!=\!T_{\rm dust}\!=\!10$\,K, density $n_{\rm\langle
  H\rangle}\!=\!10^4$\,cm$^{-3}$, extinction $A_V\!=\!10$, dust
particle density $n_{\rm dust}\!=\!1.8\times 10^{-8}$\,cm$^{-3}$ and
dust size $a\!=\!0.1\,\mu$m.  We assume large gas column densities in
order to switch off the X-rays and have large molecular shielding
factors. The integration time is chosen to be $1.7\times 10^5$\,yrs,
the assumed lifetime of TMC-1 according to \cite{McElroy2013}. We note
that these parameters are debated in the literature, see
e.g.\ \cite{Quan2008,Hincelin2011}, where the resulting abundance of
O$_2$ is crucial, because O$_2$ is not detected (observed
concentration is $<\!8\times10^{-8}$ in TMC-1
\cite{McElroy2013}). Very recently, however, \cite{Yildiz2013}
reported on a $4.5\sigma$ detection of O$_2$ from the molecular cloud
surrounding the deeply embedded low-mass class-0 protostar NGC
1333-IRAS 4A.  For the initial atomic abundances, we adopt the values
from \cite{McElroy2013}, see their Table~3.


\begin{table}[!t]
\centering
\caption{Assumed atomic and calculated abundances for dense core
  conditions. The latter are taken as initial values for the disk
  simulations in modeling stage 2.  Numbers are particle
  concentrations with respect to hydrogen nuclei, ``\#'' denote ice
  species, notation $x(-y)$ means $x\times 10^{-y}$. We only list a
  few species here, that are either abundant in the initial atomic
  (column 'atomic') or in the resultant chemical state.}
\label{tab:DenseCore}
\resizebox{15cm}{!}{
\begin{tabular}{cc}
\begin{minipage}{8cm}
{\ }\\*[-0mm]
\begin{tabular}{|c|c|cc|}
\hline
         & atomic   & UMIST\,2012$^\star$ & OSU\,2010$^\star$\\
\hline 
\hline 
H        & 5(-5)    & 2.6(-4)     & 2.2(-4)\\
H$_2$    & 0.5      & 0.5         & 0.5\\
\hline
He       & 0.09     & 0.09        & 0.09\\
\hline
C$^+$    & 1.4(-4)  & 3.0(-8)     & 1.5(-8)\\
CO       & 0        & 5.9(-5)     & 5.1(-5)\\
C        & 0        & 4.1(-5)     & 4.0(-5)\\
C\#      & 0        & 2.2(-5)     & 2.3(-5)\\
CO\#     & 0        & 1.1(-5)     & 8.1(-6)\\
\hline
N        & 7.5(-5)  & 4.9(-5)     & 4.3(-5)\\
N\#      & 0        & 1.7(-5)     & 1.6(-5)\\
N$_2$    & 0        & 3.9(-6)     & 7.0(-6)\\
N$_2$\#  & 0        & 5.8(-7)     & 8.9(-7)\\
HCN      & 0        & 1.1(-7)     & 1.4(-7)\\
HNC      & 0        & 1.0(-7)     & 1.2(-7)\\
\hline
O        & 3.2(-4)  & 1.8(-4)     & 1.9(-4)\\
O\#      & 0        & 4.1(-5)     & 4.6(-6)\\
OH\#     & 0        & 1.9(-5)     & 1.8(-5)\\
H$_2$O\# & 0        & 1.1(-5)     & 7.6(-6)\\
H$_2$O   & 0        & 3.0(-7)     & 7.6(-7)\\
O$_2$    & 0        & 1.3(-8)     & 1.2(-8)\\
\hline
\end{tabular} 
\end{minipage}
&
\begin{minipage}{8cm}
{\ }\\*[-0mm]
\begin{tabular}{|c|c|cc|}
\hline
         & atomic   & UMIST\,2012$^\star$ & OSU\,2010$^\star$\\
\hline 
\hline 
S$^+$    & 8(-8)    & 7.4(-10)     & 2.4(-9)\\
S        & 0        & 6.1(-8)      & 5.9(-8)\\
S\#      & 0        & 1.6(-8)      & 1.2(-8)\\
CS       & 0        & 2.0(-9)      & 4.8(-9)\\
CS\#     & 0        & 6.1(-10)     & 1.9(-9)\\
\hline
Si$^+$   & 8(-9)    & 1.0(-9)      & 7.6(-11)\\
Si       & 0        & 3.9(-9)      & 5.7(-9)\\
SiO      & 0        & 1.9(-9)      & 6.6(-10)\\
Si\#     & 0        & 8.9(-10)     & 1.5(-9)\\
SiO\#    & 0        & 3.3(-10)     & 1.3(-10)\\
\hline
Mg$^+$   & 7(-9)    & 5.2(-9)      & 4.9(-9)\\
Mg       & 0        & 1.5(-9)      & 1.7(-9)\\
Mg\#     & 0        & 3.6(-10)     & 4.1(-10)\\
\hline 
Fe$^+$   & 3(-9)    & 2.3(-9)      & 2.2(-9)\\
Fe       & 0        & 5.9(-10)     & 7.5(-10)\\
Fe\#     & 0        & 7.8(-11)     & 8.3(-11)\\
\hline
Ne       & 6.9(-5)  & 6.9(-5)      & 6.9(-5)\\
\hline
Ar       & 1.5(-6)  & 1.5(-6)      & 1.5(-6)\\
\hline
PAH      & 2.8(-9)  & 8.1(-10)     & 6.4(-10)\\
PAH$^-$  & 0        & 2.0(-9)      & 2.1(-9)\\
\hline
\end{tabular}
\end{minipage}
\end{tabular}}\\[2mm]
\centerline{$^\star$: \footnotesize reactions among our selection of
  species, and combined with other reactions, see text.}
\end{table}

The results of the dark core simulations are summarized in
Table~\ref{tab:DenseCore}. We get an oxygen-rich gas where oxygen and
nitrogen are mostly present in form of neutral atoms, and carbon in
form of CO, with smaller quantities already frozen out as O\#, OH\#,
H$_2$O\#, C\#, CO\#, N\# and N$_2$\# ices.  The low density and short
lifetime assumed for TMC-1 avoid large concentrations of O$_2$, and
favor the formation of simple (e.g.~atomic) ices, which have abundant
atomic/molecular counterparts in the gas phase, according to our
simple ice chemistry. However, it must be noted that if we 
integrated our reaction network for slightly longer times, or higher
densities, the O$_2$ concentration would increase rapidly, as is true in
the original, pure gas-phase, UMIST-2012 network
\cite[see][]{McElroy2013}. Adding all gaseous concentrations together,
\begin{equation}
  \langle{\rm O_{gas}}\rangle = \sum_{i\,\rm(only\ gas)} 
             \hspace*{-3mm} s^i_{\rm O} \frac{n_i}{n_{\langle\rm H\rangle}} \ ,
\end{equation} 
where, for example, $s^i_{\rm O}$ is the stoichiometric coefficient of
oxygen in molecule $i$, we get $\epsilon_{\rm
  O}\!=\!12+\log_{10}\langle{\rm O_{gas}}\rangle\!=\!8.380\,(8.381)$,
$\epsilon_{\rm C}\!=\!8.030\,(8.037)$, and $\epsilon_{\rm
  N}\!=\!7.754\,(7.755)$, where the numbers in brackets refer to the
model with the OSU rates.  These values are close to the atomic
abundances assumed in the first place, $\epsilon_{\rm O}\!=\!8.505$,
$\epsilon_{\rm C}\!=\!8.146$ and $\epsilon_{\rm N}\!=\!7.875$.

\newpage
\subsection{Stage 2: the disk simulations}

\begin{table}[!b]
\caption{Parameters of the T\,Tauri type, class-II protoplanetary disk model}
\label{tab:disk}
\begin{center}
\vspace*{-6.5mm}
\resizebox{!}{7.5cm}{
\begin{tabular}{l|c|c}
\hline
 quantity & symbol & value\\
\hline 
\hline 
stellar mass                       & $M_{\star}$   & $0.7\,M_\odot$\\
stellar luminosity                 & $L_{\star}$   & $1.0\,L_\odot$\\
effective temperature              & $T_{\rm eff}$ & $4000\,$K\\
UV luminosity                      & $L_{\rm UV}$  & $0.01\,L_\odot$\\
X-ray luminosity                   & $L_{\rm X}$   & $10^{30}\,$erg/s\\
\hline
&&\\[-2.2ex]
minimum dust particle radius       & $a_{\rm min}$  & $0.05\,\mu$m\\
maximum dust particle radius       & $a_{\rm max}$  & $3\,$mm\\
dust size dist.\ power index       & $a_{\rm pow}$  & 3.5\\
dust settling turbulence parameter & $\alpha$      & 0.001\\
max.~hollow-sphere volume ratio    & $V_{\rm max,HS}$ & 0.8\\
dust composition                   & $\rm Mg_{0.7}Fe_{0.3}SiO_3$ & 60\%\\
(volume fractions)                 & amorph.\,carbon & 15\%\\
                                   & vacuum          & 25\%\\
\hline
&&\\[-2.2ex]
disk gas mass                      & $M_{\rm gas}$ &  $3\times10^{-2}\,M_\odot$\\
disk dust mass                     & $M_{\rm dust}$ & $3\times10^{-4}\,M_\odot$\\
inner disk radius                  & $R_{\rm in}$  & 0.07\,AU\\
tapering-off radius                & $R_{\rm tap}$ & 50\,AU\\
outer disk radius                  & $R_{\rm out}$ & 200\,AU\\
column density power index         & $\epsilon$   & 1.0\\
reference scale height             & $H_0$        & 10\,AU\\
reference radius                   & $r_0$        & 100\,AU\\
flaring power index                & $\beta$      & 1.12\\ 
\hline 
&&\\[-2.2ex]
cosmic ray ionization rate & $\zeta_{\rm CR}$ &
$1.3\times10^{-17}\rm\,s^{-1\ (\star)}$\\
PAH abundance rel. to ISM          & $f_{\rm PAH}    $    & 0.01\\
chemical heating efficiency        & $\gamma^{\rm chem}$  & 0.2\\
\hline
\end{tabular}}
\end{center}
{\ }\\[-5.5mm]
\centerline{\footnotesize $^{(\star)}$: Standard value according
to \cite{Harada2010,McElroy2013}.}
\end{table}

\noindent We consider an example class-II protoplanetary disk with
parameters as listed in Table~\ref{tab:disk}.  The parameters have
been carefully chosen to match various continuum and line observations
of class-II T\,Tauri stars concerning SED shape (clearly
visible 10\,$\mu$m and 20\,$\mu$m silicate emission features,
decreasing SED-slope beyond 20\,$\mu$m, typical for non-transitional
disks), near-IR excess ($0.15\rm\,L_\odot$ between 2$\,\mu$m and
7$\,\mu$m), mm-flux (130\,mJy at 140\,pc), mm-slope
$\beta=-\Delta\log(F_\nu)/\Delta\log(\lambda)-2\!=\!0.3$,
[OI]\,63\,$\mu$m line flux ($4\times 10^{-17}\rm\,W/m^2$ at 140\,pc)
and various CO sub-mm line fluxes. The stellar parameters describe a
T\,Tauri star of spectral type K7 with an age of about 1.6 Myrs. The
disk extends radially from 0.07\,AU to 200\,AU, and has an assumed
radial surface density profile as
$\Sigma(r)\!\propto\!r^{-\epsilon}\exp\big(\!\!-\!\frac{r}{R_{\rm tap}}\big)$.
The resulting densities in the midplane are
$n\!\approx\!(10^{16}-10^6)$ cm$^{-3}$, depending on $r$, and the
midplane temperatures are as low as $T\!\approx\!(300-5)$\,K,
 encompassing radii $r\!=\!0.15\!-\!200$\,AU (disregarding the hot inner
rim with temperature $\approx\!1500\,$K here). Dust settling is
included according to \cite{Dubrulle1995}, assuming an equilibrium
between upward turbulent mixing and downward gravitational settling,
which leads to higher dust/gas ratios in the midplane, in particular
in the outer midplane. The midplane regions are entirely shielded from
the stellar UV and even from the stellar X-rays. The vertical
extinction of the midplane is $A_V\!\approx\!1500$ at 1\,AU and still
$A_V\!\approx\!15$ at $r\!=\!50\,$AU. The radial $A_V$ are much 
larger.

The above described conditions are typical for the midplane only, and
we will entirely focus on the central midplane results in this paper, where
planet formation occurs. Since stellar UV and X-rays can penetrate the
upper, thinner disk layers, these layers have a very
different chemistry. The upper layers are much warmer, and the high-energy
photons drive an active X-ray/photo-chemistry producing most of the
observable line emissions. In these layers, the chemical relaxation
timescales are short, and the application of kinetic chemical
equilibrium is justified \cite{woi09}. Therefore, it would be an error
to generalize the chemical midplane results as described in this paper
to the entire disk, or to the line-emitting regions.

We use the dense core concentrations from Table~\ref{tab:DenseCore} as
initial values, reset the clock, and integrate forward our chemical
rate network in disk configuration, using a well-iterated spatial
density, temperature and radiation field structure of the disk
obtained before with a standard model, where kinetic chemical
equilibrium is assumed. The physical conditions in the 2D disk are
hence kept constant in time during modeling stage 2.

\begin{figure}[!t]
\centering
\vspace*{-4mm}
\hspace*{-2mm}\begin{tabular}{cc}
$\bf t\!=\!1\,$Myr & $\bf t\!=\!3\,$Myrs\\[-1mm]
\includegraphics[trim=10mm 21mm 17mm 25mm,clip,width=87mm]
                {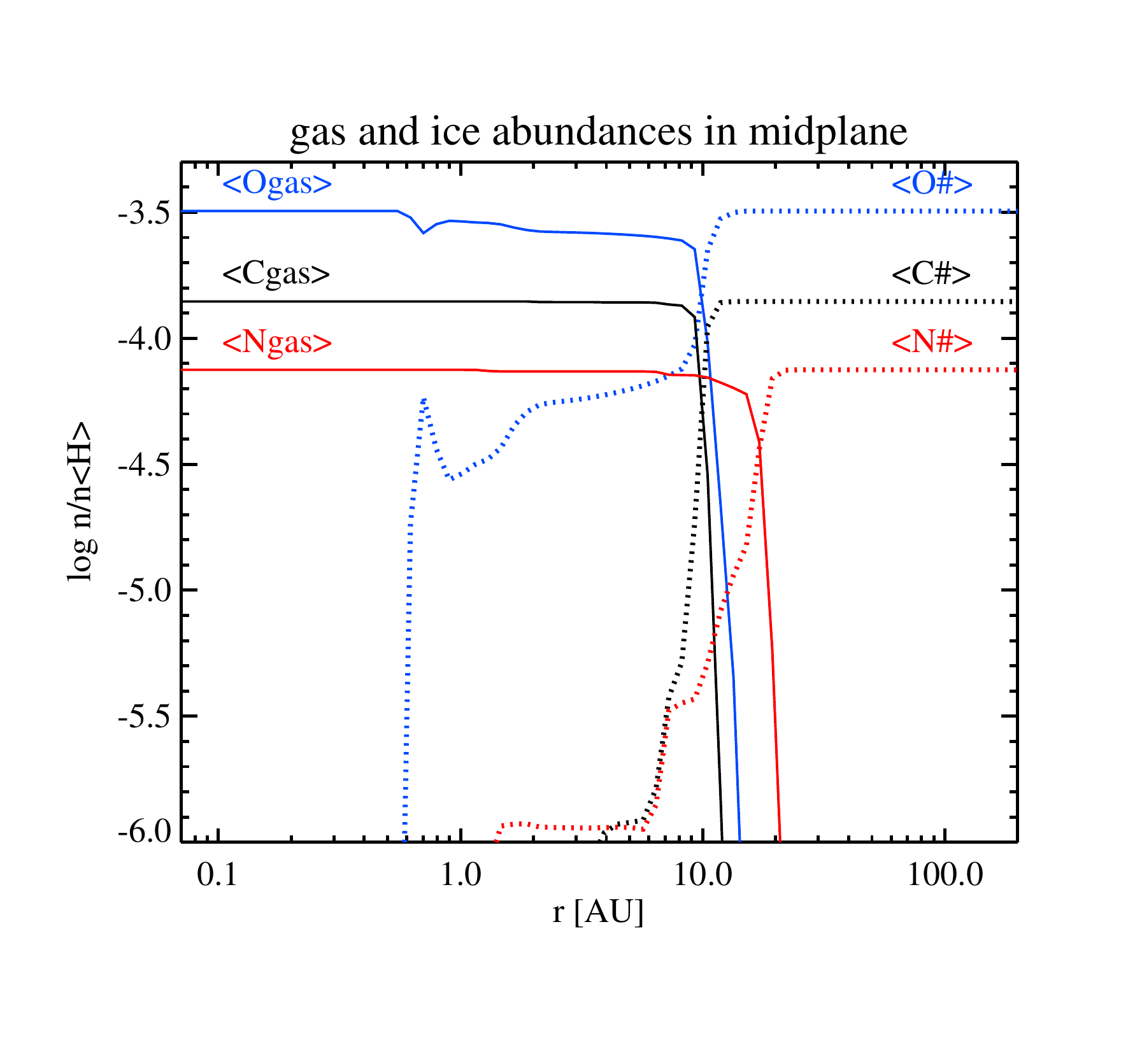} &
\hspace*{-5mm}
\includegraphics[trim=10mm 21mm 17mm 25mm,clip,width=87mm]
                {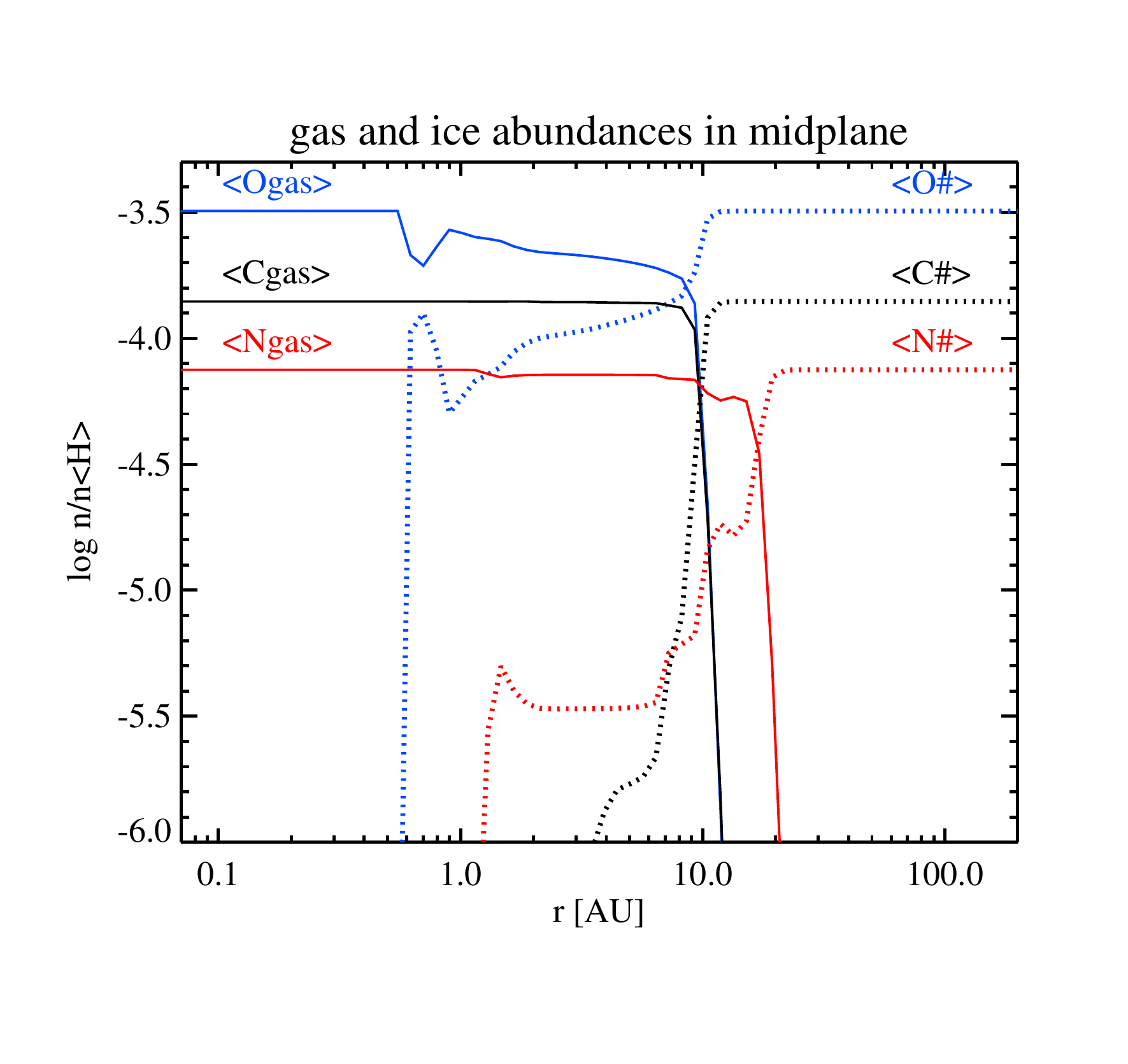} \\[-2mm]
$\bf t\!=\!10\,$Myrs & $\bf t\!=\!\infty$\\[-1mm]
\includegraphics[trim=10mm 21mm 17mm 25mm,clip,width=87mm]
                {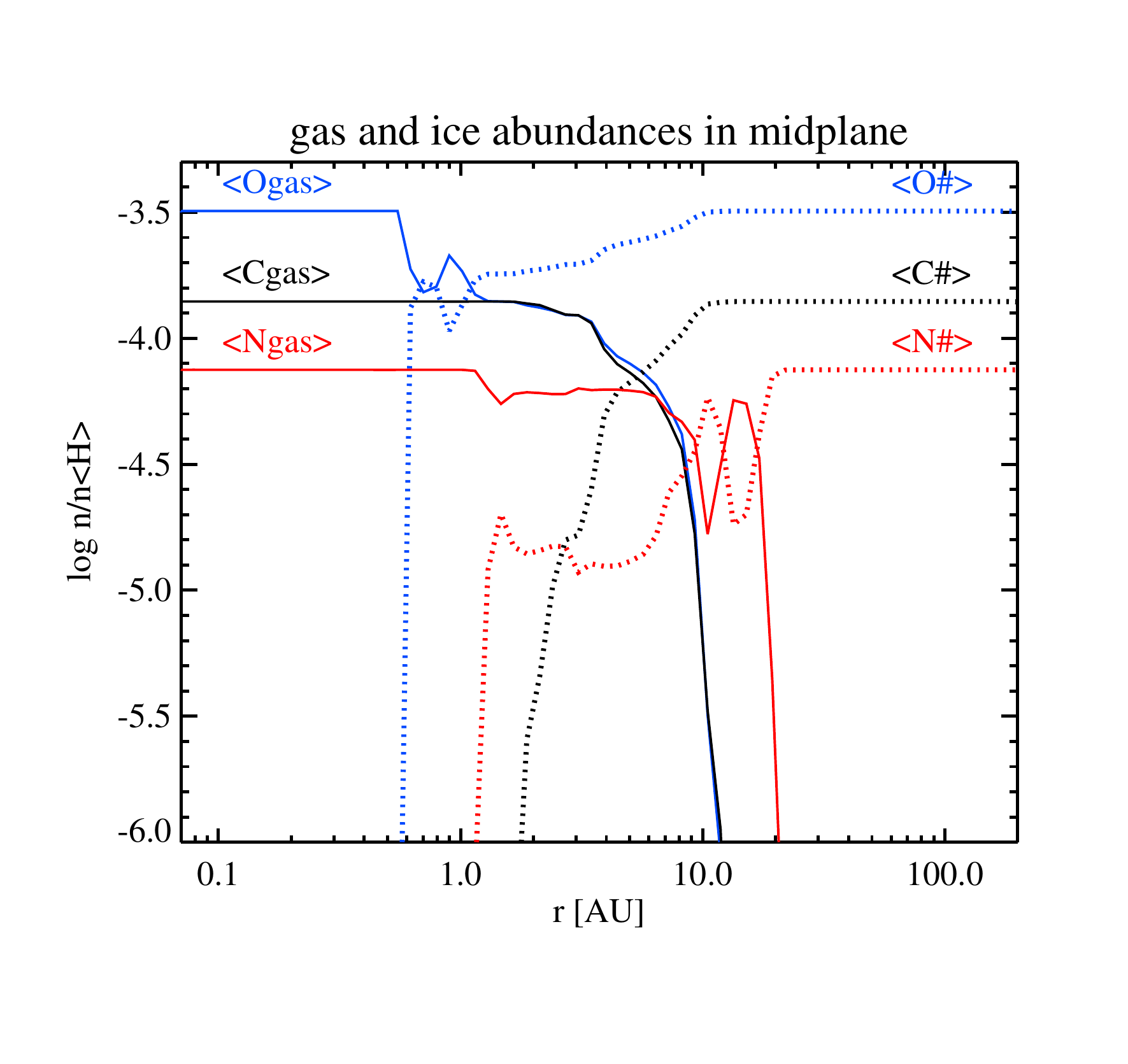} &
\hspace*{-5mm}
\includegraphics[trim=10mm 21mm 17mm 25mm,clip,width=87mm]
                {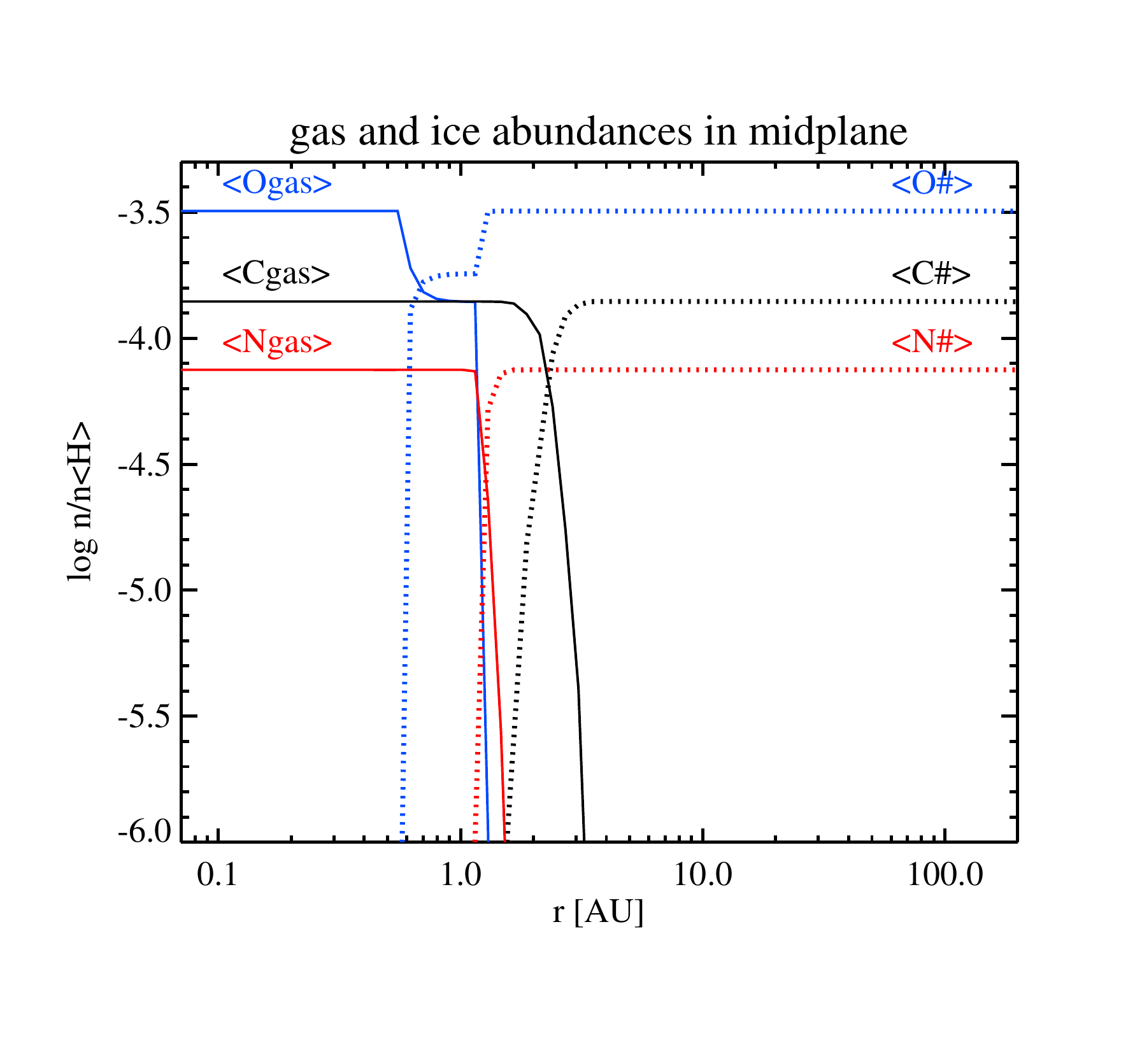} 
\end{tabular}
\caption{Time evolution of total gas (solid) and total ice (dotted)
  abundances of oxygen (blue), carbon (black), and nitrogen (red) in
  the midplane, according to a time-dependent ProDiMo model, based on
  the UMIST-2012 rates, for a T\,Tauri type protoplanetary disk. The
  y-axis shows the concentration with respect to hydrogen nuclei, take
  this value $+12$ to get the usual element abundances $\epsilon$
  (i.e.\ $\epsilon_H\!=\!12$).}
\label{fig:epsDisk_UMIST2012}
\end{figure}

\begin{figure}[!t]
\centering
\vspace*{-4mm}
\hspace*{-2mm}\begin{tabular}{cc}
$\bf t\!=\!1\,$Myr & $\bf t\!=\!3\,$Myrs\\[-1mm]
\includegraphics[trim=10mm 21mm 17mm 25mm,clip,width=87mm]
                {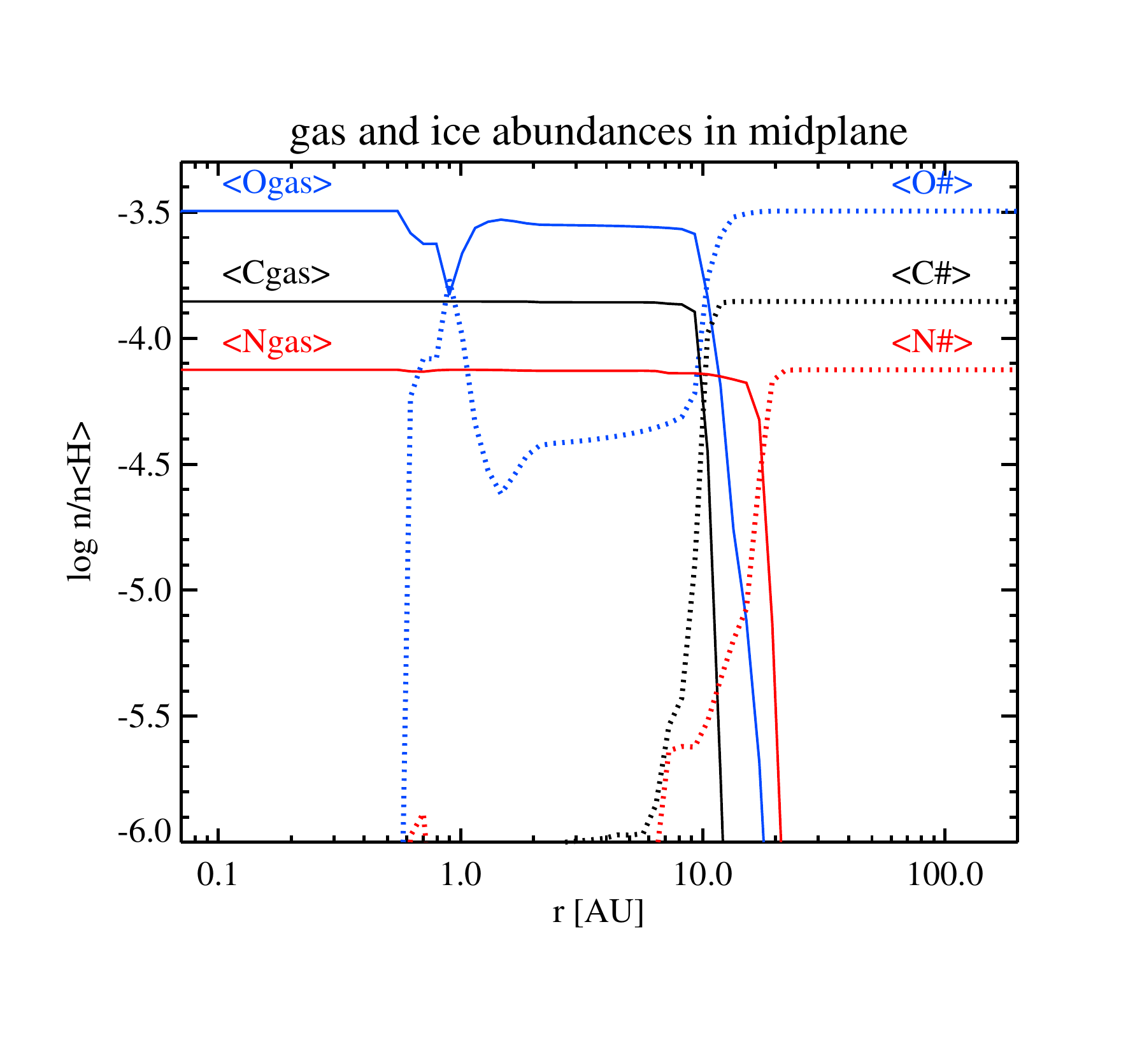} &
\hspace*{-5mm}
\includegraphics[trim=10mm 21mm 17mm 25mm,clip,width=87mm]
                {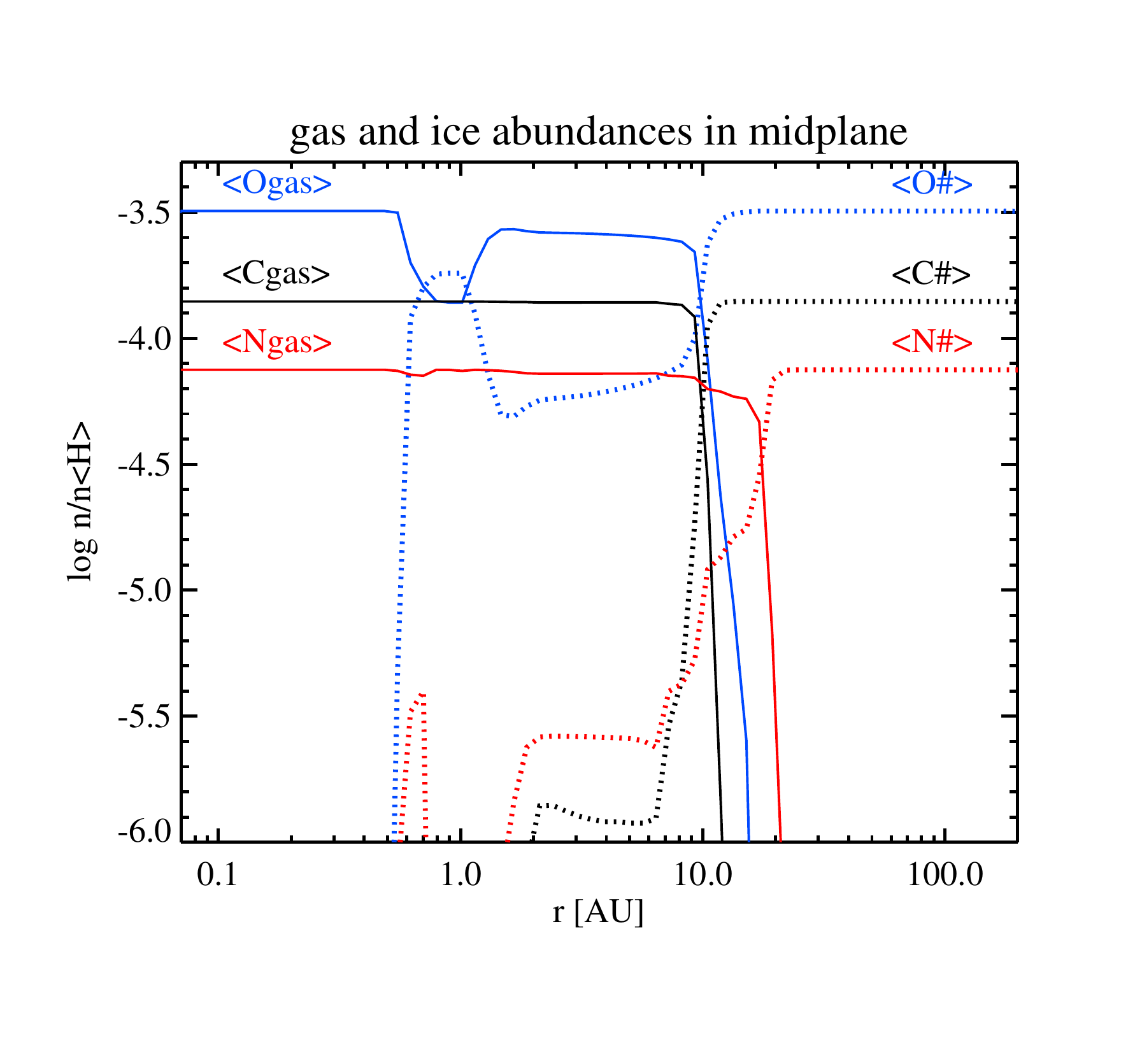} \\[-2mm]
$\bf t\!=\!10\,$Myrs & $\bf t\!=\!\infty$\\[-1mm]
\includegraphics[trim=10mm 21mm 17mm 25mm,clip,width=87mm]
                {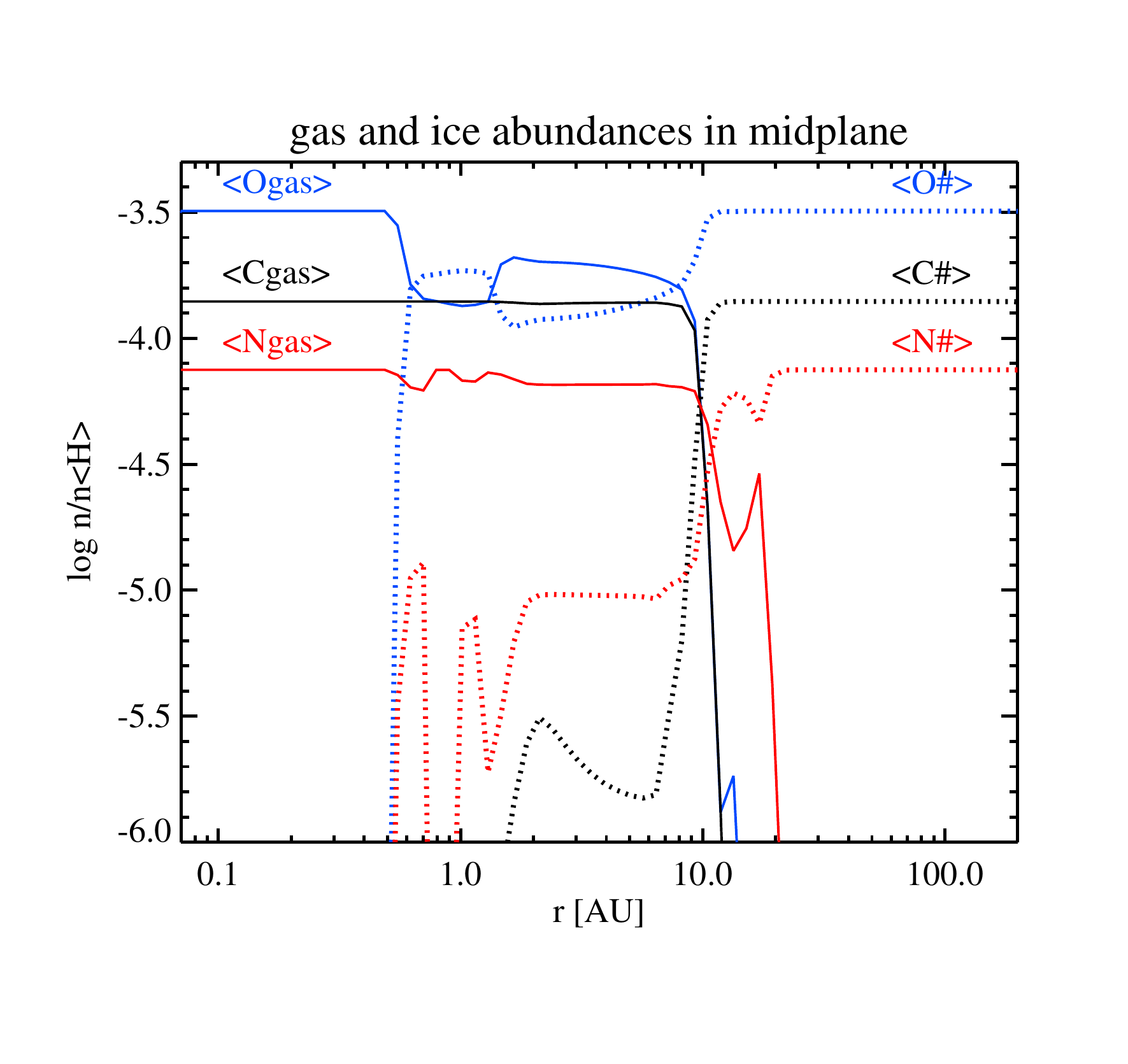} &
\hspace*{-5mm}
\includegraphics[trim=10mm 21mm 17mm 25mm,clip,width=87mm]
                {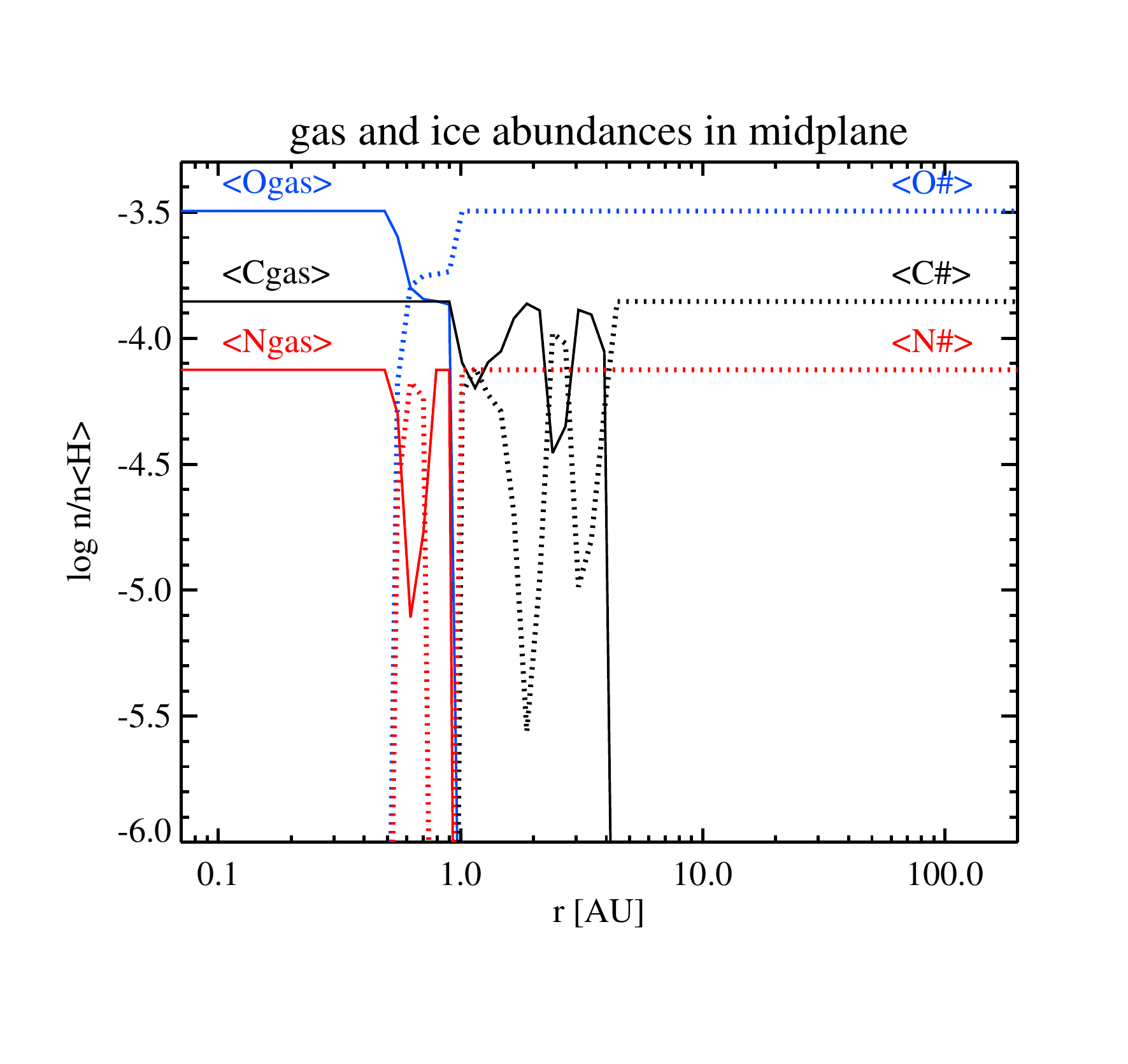} 
\end{tabular}
\caption{Same as figure~\ref{fig:epsDisk_UMIST2012}, but for the 
  ProDiMo model based on the OSU-2010 rates.}
\label{fig:epsDisk_OSU2010}
\end{figure}

\subsection{Results}\label{ss:results}

\noindent The time-dependent segregation of carbon, nitrogen and
oxygen into gas and ice in the midplane is shown in
Figs.~\ref{fig:epsDisk_UMIST2012} and \ref{fig:epsDisk_OSU2010}, based
on the models with the UMIST-2012 and the OSU-2010 rates,
respectively.  According to the model, the midplane is chemically
subdivided into three different radial zones, separated by the H$_2$O
and CO ice-lines. The innermost zone ($r\!<\!0.6\,$AU, $T\!<\!150\,$K,
$n_{\rm\langle H\rangle}\!>\!5\times 10^{14}\rm\,cm^{-3}$) is too hot
for any stable ices, hence the gas abundances are equal to the assumed
total element abundances. The sum of gas and ice abundances is
  prescribed by the ``element abundances'' in the model. Abundant
molecules here are H$_2$O, CO, CO$_2$, HCN, HNC, CH$_4$ and NH$_3$,
similar to the ultra-cool atmospheres of brown dwarfs or giant gas
planets (compare Fig.~\ref{fig:130030_gg}).

In the outer zone, beyond the CO ice-line ($r\!>\!13\,$AU,
$T\!<\!25\,$K, $n_{\rm\langle H\rangle}\!<\!5\times
10^{11}\rm\,cm^{-3}$), oxygen and carbon are quickly converted to ices
(mainly H$_2$O\# and CO\#). Nitrogen freezes out in form of N\# and
N$_2$\# a bit further out ($r\!>\!20\,$AU), because of the slightly
lower adsorption energies. These statements are valid already after
some 100 yrs. The initial freeze-out of condensable molecules is
actually a very fast process (see Table~\ref{tab:eps}). The adsorption
timescale $\tau_{\rm ads}$, i.e.\ the timescale for a molecule to hit
and stick to the surface of a grain is
\begin{equation}
  \tau_{\rm ads}^{-1} = \alpha\,n_{\rm dust}\,4\pi\langle a^2\rangle\,
                       v_{\rm th}\ ,
\end{equation}
where $\alpha\!\approx\!1$ is the sticking coefficient, $n_{\rm
  dust}\rm\,\rm[cm^{-3}]$ the local dust particle density, $\langle
a^2\rangle$ the mean of the squared dust particle radii, averaged over
the size distribution, and $v_{\rm th}\!=\!\sqrt{kT/(2\pi\,m)}$ the
thermal velocity of a molecule with mass $m$. In the disk model, for a
molecule like water ($m\!=\!18\,$amu), this timescale is as short as
$\rm 1\,$sec at the inner rim, and $100\,$yrs at the outer radius,
despite the tapering-off surface density assumed.  What turns the ice
formation actually into a slow process is the chemical conversion of
the gas phase into condensable molecules prior to freeze-out.

Thus, already after 1\,Myrs, the distant midplane gas contains
practically no molecules other than H$_2$. The atmosphere of a Uranus
or Neptune-like planets, if composed of such gas alone, would contain
practically no oxygen, carbon or nitrogen.

\begin{figure}[t]
\centering
\vspace*{-2mm}
\hspace*{0mm}\includegraphics[trim=0mm 77mm 0mm 58mm,clip,width=180mm]
                {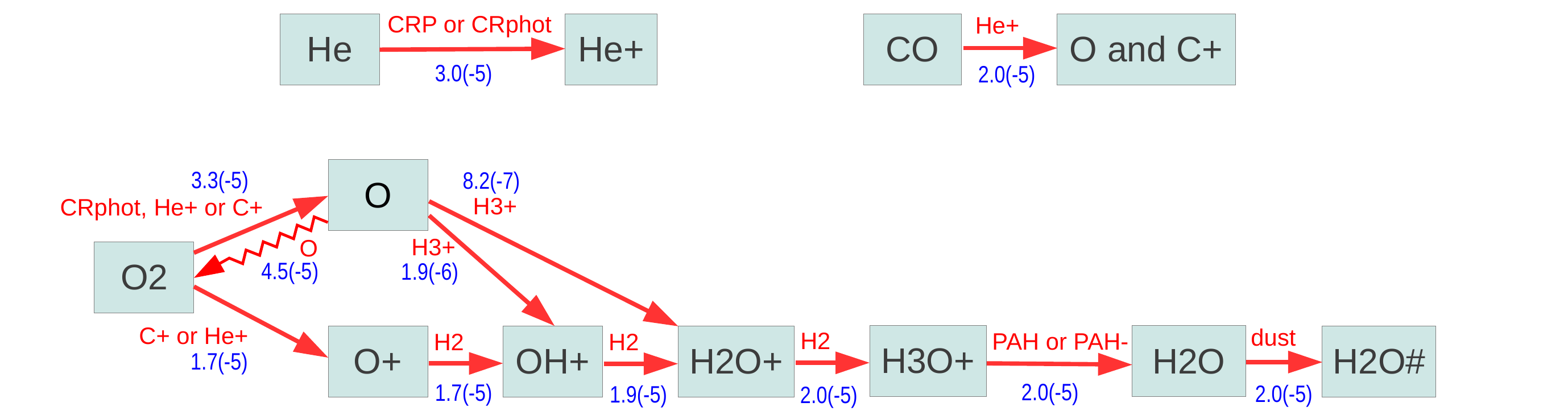}
\caption{Destruction of molecular oxygen and formation of water ice,
  after 1\,Myr in the midplane of the UMIST-2012 model at 2.5\,AU,
  where temperature and density are 75\,K and $3\times
  10^{13}\rm\,cm^{-3}$, respectively. ``\#'' denotes ice species.  The
  red species indicate the reaction partners, and the blue numbers are
  the reaction rates in $\rm s^{-1}cm^{-3}$.}
\label{fig:O2}
\end{figure}

In the sandwich zone between the H$_2$O and the CO ice-lines, that is
between 0.6\,AU and 13\,AU in the particular model, the results are
time-dependent. The earlier epochs ($t\!\lesssim\!3$\,Myrs) are
characterized by the slow build up of additional H$_2$O\# from the
abundant gas phase molecules CO, CO$_2$, O$_2$, N and N$_2$. The slow
formation of water ice is what makes our results differ from
\cite{Oberg2011}. This is because of the initial formation of O$_2$
parallel to H$_2$O, and the temperatures being too warm for O$_2$ to freeze
out.  This way, most of the gaseous oxygen in the midplane is soon
locked into the stable and chemically inert O$_2$.  In order to form
additional H$_2$O\# under those circumstances, the O$_2$ must be
dissociated first, and this proceeds in the model either by cosmic
ray ionisations or by reactions with C$^+$, which are both slow processes.

Figure~\ref{fig:O2} shows some details of the chemical paths involved.
Initially, a helium atom is ionized by cosmic rays, and the He$^+$
collides with a CO molecule to dissociate it into C$^+$ and O. The
C$^+$ then attacks O$_2$.  If O$_2$ is dissociated into neutral O
atoms, quick radiative association reactions with other O atoms will
re-form O$_2$.  However, when O$^+$ is produced, there is a quick
linear reaction chain which stepwise adds hydrogen atoms via reactions
with H$_2$ to form H$_3$O$^+$, and H$_3$O$^+$ then recombines at the
surface of PAH molecules to form H$_2$O, which then freezes out.
Simplistically speaking, for every CO molecule un-blocked by cosmic
rays, there will be one O$_2$ molecule destroyed, and one new H$_2$O\# ice
unit formed.  The effective destruction timescale for O$_2$, in the
example shown, is
\begin{equation}
  \tau_{\rm O_2} = \frac{n_{\rm O_2}}{dn_{\rm O_2}/dt}
                = \frac{1.6\times10^9\rm\,cm^{-3}}
                       {2.0\times10^{-5}\rm\,s^{-1}cm^{-3}}
                \;\approx\; 2.5\rm\,Myrs
\end{equation} 
in the UMIST-2012 model, and $\approx\!6\,$Myrs in the OSU-2010 model.
The difference between the two models can be traced back to the CR
induced secondary UV reaction $\rm He + CRphot \to He^+ + e^-$, in
addition to the primary reaction $\rm He + CRP \to He^+ + e^-$. The
secondary reaction, which about doubles the ionization rate of He,
seems new in UMIST-2012 (compared to UMIST-2006), and was apparently
not incorporated into the OSU-2010 reaction rates either.
Consequently, our OSU-2010 model has about a factor of two less
He$^+$, and accordingly less C$^+$, both required to form ionized
oxygen, the main precursor of water.

As O$_2$ is slowly consumed and converted into H$_2$O\#, the gaseous
C/O-ratio increases and reaches unity after about 5\,Myrs, but then C/O does {\it not} increase further. All
  timescales mentioned in the remainder of the text belong to the
  UMIST-2012 model, and scale with the assumed cosmic ray ionization
  rate. Although CO is
continuously dissociated by cosmic rays (timescale about
$\sim\,$6\,Myrs in the UMIST-2012 model, $\sim\,$14\,Myrs in the
OSU-2010 model), the liberated O rather reacts with other molecules
like CS and CN to reform CO. Some tiny amounts of H$_2$O\# do actually
form via OH, but the associated timescale to convert CO into H$_2$O\#,
in case $\rm C/O\!\approx\!1$, is huge, larger than 100\,Myrs.

\begin{table}[!t]
\centering
\caption{Carbon, nitrogen and oxygen gas abundances at selected times
  and locations in the disk, according to disk models using the
  UMIST-2012 and OSU-2010 chemical rates.}
\vspace*{-2mm}
\resizebox{12cm}{!}{\begin{tabular}{|c|ccccc|ccccc|}
\hline
              &\multicolumn{5}{|c|}{UMIST-2012}
              &\multicolumn{5}{|c|}{OSU-2010}\\
\cline{2-11}
              & 0.3\,AU & 1\,AU & 3\,AU & 10\,AU & 30\,AU 
              & 0.3\,AU & 1\,AU & 3\,AU & 10\,AU & 30\,AU \\
\hline
\multicolumn{11}{|c|}{$t\!=\!0$}\\
\hline
$\epsilon$(O) & 8.38 & 8.38 & 8.38 & 8.38 & 8.38 
              & 8.38 & 8.38 & 8.38 & 8.38 & 8.38 \\
$\epsilon$(C) & 8.03 & 8.03 & 8.03 & 8.03 & 8.03
              & 8.04 & 8.04 & 8.04 & 8.04 & 8.04 \\
$\epsilon$(N) & 7.75 & 7.75 & 7.75 & 7.75 & 7.75 
              & 7.75 & 7.75 & 7.75 & 7.75 & 7.75 \\
\hline
\multicolumn{11}{|c|}{$t\!=\!100$\,yrs}\\
\hline
$\epsilon$(O) & 8.51 & 8.49 & 8.46 & 8.35 & 2.41
              & 8.51 & 8.49 & 8.47 & 8.36 & 4.47 \\
$\epsilon$(C) & 8.15 & 8.15 & 8.14 & 7.83 & 5.56
              & 8.15 & 8.15 & 8.14 & 7.84 & 4.63 \\
$\epsilon$(N) & 7.88 & 7.87 & 7.87 & 7.86 & 5.03
              & 7.88 & 7.88 & 7.87 & 7.87 & 2.55 \\
\hline
\multicolumn{11}{|c|}{$t\!=\!1$\,Myr}\\
\hline
$\epsilon$(O) & 8.51 & 8.46 & 8.42 & 8.12 & 1.37
              & 8.51 & 8.32 & 8.45 & 8.26 & 1.43 \\
$\epsilon$(C) & 8.15 & 8.15 & 8.14 & 7.70 & 1.24
              & 8.15 & 8.15 & 8.14 & 7.76 & 1.33 \\
$\epsilon$(N) & 7.88 & 7.88 & 7.87 & 7.85 & 1.75
              & 7.88 & 7.88 & 7.87 & 7.86 & 1.95 \\
\hline
\multicolumn{11}{|c|}{$t\!=\!3$\,Myrs}\\
\hline
$\epsilon$(O) & 8.51 & 8.42 & 8.33 & 7.65 & 0.24
              & 8.51 & 8.14 & 8.42 & 8.08 & 0.37 \\
$\epsilon$(C) & 8.15 & 8.15 & 8.14 & 7.57 & 0.83
              & 8.15 & 8.15 & 8.14 & 7.68 & 0.93 \\
$\epsilon$(N) & 7.88 & 7.88 & 7.85 & 7.80 & 1.77
              & 7.88 & 7.87 & 7.86 & 7.82 & 2.04 \\
\hline
\multicolumn{11}{|c|}{$t\!=\!10$\,Myrs}\\
\hline
$\epsilon$(O) & 8.51 & 8.27 & 8.09 & 6.80 &-1.53
              & 8.51 & 8.13 & 8.30 & 7.61 &-1.53 \\
$\epsilon$(C) & 8.15 & 8.15 & 8.09 & 6.79 &-0.25
              & 8.15 & 8.15 & 8.14 & 7.60 &-0.12 \\
$\epsilon$(N) & 7.88 & 7.87 & 7.80 & 7.37 & 1.73
              & 7.88 & 7.84 & 7.82 & 7.71 & 2.02 \\
\hline
\multicolumn{11}{|c|}{$t\!=\!30$\,Myrs}\\
\hline
$\epsilon$(O) & 8.51 & 8.15 & 7.29 & 1.83 &-1.53
              & 8.51 & 8.13 & 8.06 & 7.18 &-1.54 \\
$\epsilon$(C) & 8.15 & 8.15 & 7.30 & 2.09 &-0.39
              & 8.15 & 8.15 & 8.06 & 7.19 &-0.26 \\
$\epsilon$(N) & 7.88 & 7.87 & 6.78 & 1.64 & 1.68
              & 7.88 & 7.83 & 7.61 & 6.82 & 1.96 \\
\hline
\multicolumn{11}{|c|}{$t\!=\!\infty$}\\
\hline
$\epsilon$(O) & 8.51 & 8.15 & -7.62 & -24.8 & -13.8
              & 8.51 & 5.12 & -5.18 & -24.8 & -13.8\\
$\epsilon$(C) & 8.15 & 8.15 &  6.73 & -3.23 & -2.38
              & 8.15 & 7.93 &  8.03 & -2.95 & -2.02\\
$\epsilon$(N) & 7.88 & 7.87 & -4.92 & -29.3 & -14.6
              & 7.88 & 1.92 & -12.9 & -29.6 & -14.5\\
\hline
\end{tabular}}
\label{tab:eps}
\end{table}

At later epochs ($t\!\gtrsim\!10$\,Myrs) additional, very stable ices
with rare gaseous counterparts are formed, in particular NH$_2$\#,
NH$_3$\#, C$_2$H$_2$\#, CH$_3$OH\#, CH$_3$\# and CH$_4$\#.  In fact,
the simple ices formed in the first place are now slowly converted
into these ``late ices'' \cite[see
  also][]{Chaparro2012a,Chaparro2012b}.  Since our model has only very
limited surface chemistry \cite{Kamp2013}, this conversion requires to
temporarily evaporate the simple ices (by cosmic ray desorption), to
convert the respective molecules in the gas phase into different ones
by cosmic-ray chemistry, and then to freeze out the new molecules.  These
processes are extremely slow.

At an age of about 10\,Myrs, considerable fractions of the late ices
have built up, and the sandwich zone starts to shrink from the outside
in. The outer part of the former sandwich zone joins the outer disk in
becoming virtually molecule-free (except for H$_2$), while the simple
ices are steadily converted into their most stable, complex forms. The
lower right plots in Figs.~\ref{fig:epsDisk_UMIST2012} and
\ref{fig:epsDisk_OSU2010} shows the ``fictive end stadium'',
calculated in kinetic chemical equilibrium, where nitrogen disappears
from the gas phase at about 1.5\,AU, triggered by NH$_3$ condensation,
and gaseous carbon disappears at $r\!\gtrsim\!3$\,AU, due to the
condensation of C$_2$H$_2$ and CH$_3$OH ices. The chemical equilibrium
model is also featured by $\rm C/O\!\gg\!1$ outside of about 1.5\,AU
where the remaining gas is devoid of CO. Instead, organic molecules
are abundant, in form of small hydro-carbon chain molecules like
C$_2$, C$_3$, C$_3$H, C$_2$H$_2$ and C$_3$H$_2$.

Because all chemical processes described above are driven by cosmic ray
ionization, the associated timescales are density-independent, i.e.
the whole disk will undergo the chemical conversion $\rm C/O\!\to\!1$ 
in a coherent way, between the H$_2$O and CO
ice-lines. However, the initial formation of the stable gases CO,
O$_2$, CO$_2$, N$_2$, etc., depends on local conditions and that
explains the differences at one particular time in
Figs.~\ref{fig:epsDisk_UMIST2012} and \ref{fig:epsDisk_OSU2010}.
Thus, cosmic rays provide a ``clock'' for the chemical conversion 
$\rm C/O\!\to\!1$, driven by water ice formation, in the middle sections
of protoplanetary disk midplanes (see Table~\ref{tab:eps}).

\subsection{Discussion}

\noindent In this paper, we have investigated the chemical pre-conditions
for planet formation, in terms of gas and ice abundances as function
of time and position  in the midplane of a protoplanetary disk.  
Under the dense and shielded conditions, the chemical composition in
the disk midplane soon becomes quite simple. Only the most stable,
almost inert, neutral molecules like H$_2$O, O$_2$, CO, CO$_2$,
CH$_4$, N$_2$ and NH$_3$ will soon contain the vast majority of the
elements, similar as in brown dwarf atmospheres.  At radii where 
the disk midplane is cold enough for the corresponding ice phases to be 
thermally stable, those molecules freeze out after short times ($\ll
10^3$\,yr). After that initial short period of relaxation, the
chemistry then ``comes to a halt'', meaning that almost no chemical
processes occur anymore in the disk midplane -- the chemical
timescales increase towards millions of years.

All remaining chemical activity is then entirely due to cosmic ray
(CR) hits, and we have applied a standard CR ionization rate of H$_2$
of $1.3\!\times\!10^{-17}\rm\,s^{-1}$ \cite{McElroy2013} throughout
the disk, which provides a slowly ticking clock, as with every
dissociated O$_2$ and CO molecule, there are opportunities to form
other molecules which can freeze out, like water. The selective
  freeze-out of molecules containing oxygen leads to $\rm C/O\!\to\!1$
  in the gas phase inward of the CO ice-line ($\approx\!20\,$K), on
  timescales of several Myrs. The question arises what happens if
cosmic rays do not even reach the disk, but are shielded by magnetic
fields or by inelastic collisions with the surrounding gas, see
e.g.\ \cite{Cleeves2013}.  In that case, the midplane ionization might
be dominated by the decay of $^{26}$Al, resulting in a much lower
midplane ionization rate of $4\times10^{-19}\rm s^{-1}$
\cite{Meijerink2008}.  The results of this paper are still valid then,
but with the time-axis re-scaled.  If the ionization rate is indeed as
low as $4\times10^{-19}\rm s^{-1}$, there will be  indeed no
  further chemical activity in the midplane during the lifetime of
the protoplanetary disk. If, however, the ionization rate is
  $10\times$ or $100\times$ larger, for example if the star formation
  region has produced nearby supernova explosions, which produce new
  cosmic rays locally, the conversion to $\rm C/O\!\to\!1$ might take
only several $10^5$ or $10^4$ years, respectively.

The large initial quantities of O$_2$ predicted by our disk model may
 not be a reliable result. Molecular oxygen remains undetected in
  a small number of observed interstellar clouds of varying ages
  \cite{Goldsmith2000,Bergin2000}.  These observational findings by
  the SWAS/Odin missions have triggered new world-wide efforts to
  refine the chemical rate networks, and to include additional
  complicated surface chemical processes.  Different groups have
  presented different solutions how to keep the O$_2$ concentration
  within the observed limits under dark cloud conditions
  \cite{Quan2008,Hincelin2011}.  Assuming quite young ages for the
  clouds seems to provide the easiest solution \cite{McElroy2013},
but typically results in an overabundance of gaseous H$_2$O
\cite{Wakelam2006}.  We, too, can avoid the over-prediction of O$_2$
in the dark cloud model by assuming low densities and young cloud
ages,  but we cannot avoid the formation of massive amounts of
  O$_2$ in the disk model during the first $\sim 10^5\,$yrs under the
  high-density conditions in the disk, neither with the UMIST-2012,
  nor with the OSU-2010 reaction rates. We are possibly missing some
complicated surface-chemical processes which convert O$_2$ on contact
with dust grain surfaces. 
  
 An alternative idea how to resolve the O$_2$ mystery is to
  consider discharge processes (lightning) like in substellar
  atmospheres. The degree of ionization of the disk midplane can be
  altered, at least temporarily, by small-scale or large-scale
  discharge processes between single grains or ensembles of grains
  that undergo collisional ionization due to the turbulent character
  of the midplane disk material \cite{hell2011a, hell2011b,
    hell2013}. Such processes might trigger Alfv{\'e}n ionization of
  the O$_2$ and other molecules in a weakly magnetized disk
  \cite{stark2013} potentially increasing the chemical activity in the
  disk midplane.
 
Another effect, that has been disregarded in this paper, is that large
planetesimals, especially the large parental bodies that form during
oligarchic growth,  may heat up internally due to the heating provided
by radioactive decay of $^{26}$Al \cite{Henke2012}. The question
arises then to what degree planet cores lose their ices, or even the
more volatile constituents of their refractory material, just like
comets do when they come too close to the sun, prior to the run-away gas
accretion.  These questions, however, go beyond the scope of this
paper. If the ice elements C, N and O are completely returned to the gas
phase, just prior to the rapid phase of gas accretion, the gas might
re-gain its primordial C/N/O element composition, or may even locally
exceed it, if the dust/gas ratio was locally enhanced, for example due
to gravitational settling.

\section{Cloud formation processes in extrasolar planets}\label{s:cloudmod}

\noindent 
Inspired by the complexity of the changing carbon and oxygen
abundances during the evolution of a protoplanetary disk, we present
first tests of the impact of changing oxygen and carbon abundances on
the cloud formation in ultra-cool, planetary atmospheres as one of the
essential modeling complexes.

All solar system planets with an atmospheres have clouds of various
compositions. Only our Earth has exactly the right amount of clouds to
allow vegetation to grow by letting the Sun-light pass through while
still protecting the surface from too much Sun-light, and by
transporting and releasing water over the continents.  Extrasolar
planets should also have clouds because their atmospheres are
sufficiently cool, but their composition will be very different from
what we know from Earth and the solar system; they are made of various
kinds of minerals in giant gas planets or in warm atmosphere portions of cooler planets \cite{hwt08,wit09}. Transit observations of the
exo-planet HD189733b \cite{ev13, pon13}  support these results by
 suggesting the presence of small silicate grains (haze) in the
upper layers. Sub-micron size aerosol particles were  suggested
in the upper atmosphere of WASP-12b \cite{sing13}.  To complicate
things, we cannot take cloud measurements of these exoplanets like we
do for Earth and, to some extant, for Venus and Jupiter. This
situation requires us to think carefully about the cloud formation
processes and their consistent modeling in much more detail than
needed for the solar system. The emphasis is here on the coupling and
simultaneous treatment of all possible processes as local situations
can vastly change from exo-planet to exo-planet. We will summaries the
processes that lead to and are involved in the formation of
atmospheric clouds in ultra-cool, planetary objects in the next
section, and discuss relevant results of our cloud formation model for
illustration.
A comprehensive outline of the body of equations of our kinetic
approach to cloud formation is given in \cite{hefo13}. 

\begin{figure}[htbp]
\centering \includegraphics[scale=0.8]{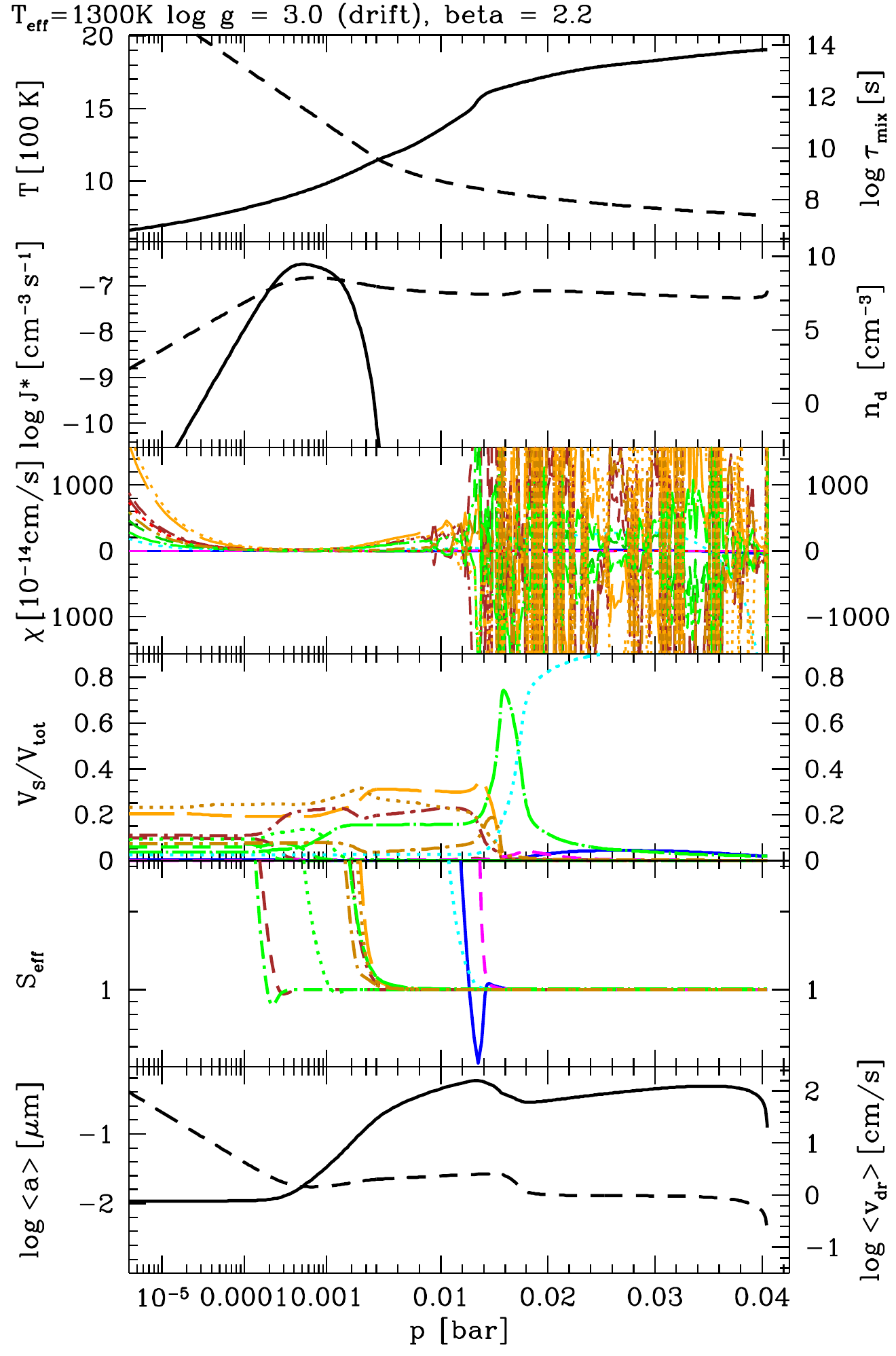}
\caption{Structure and physical properties of a mineral cloud in the
  atmosphere of a planet with T$_{\rm eff}$=1300K, log(g)=3.0 and
  solar element abundances. The model atmosphere structure is taken
  from the {\sc Drift-Phoenix} grid \cite{wit09}.  The solar {\sc
      Drift-Phoenix} grid spans T$_{\rm eff}$=1000$\ldots$3000K,
    log(g)=3.0$\ldots$6.0.\newline {\bf 1$^{\rm st}$ panel:} left - gas phase
  temperature T$_{\rm gas}$ [K], right - time scale of convective
  up-mixing $\tau_{\rm mix}$ [s]; {\bf 2$^{\rm nd}$ panel:} left - nucleation
  rate J$_*$ [cm$^{-3}$ s$^{-1}$], right - number density of dust
  particles n$_{\rm d}$ [cm$^{-3}$]; {\bf 3$^{\rm rd}$ panel:} growth velocity
  of different materials $\chi$ [cm/s]; {\bf 4$^{\rm th}$ panel:} particle
  material composition in volume fraction V/V$_{\rm s}$ ($\sum_{\rm s}
  V_{\rm s}$ - total dust volume); {\bf 5$^{\rm th}$ panel:} effective
  supersaturation ratio for each material S$_{\rm eff}$; 6$^{\rm th}$ panel:
  left - cloud particle mean size $<$a$>$ [$\mu$m], right - mean drift
  velocity v$_{\rm dr}$ [cm/s].\newline The color/ line coding is the
  same for all panels and plots: TiO$_2$[s] - solid blue,
  Mg$_2$SiO$_4$[s] - orange long-dash, MgO[s] - dark orange dot dash,
  SiO[s] - brown dost short dash, SiO$_2$[s] - brown dot dash; Fe[s] -
  green dot long dash ; Al$_2$O$_3$[s] - cyan dotted , CaTiO3$_3$[s] -
  magenta dashed. Only a subset of all 12 materials is depicted.}
\label{fig:13003.0ref}
\end{figure}

\begin{figure}[htbp]
\centering \includegraphics[scale=0.8]{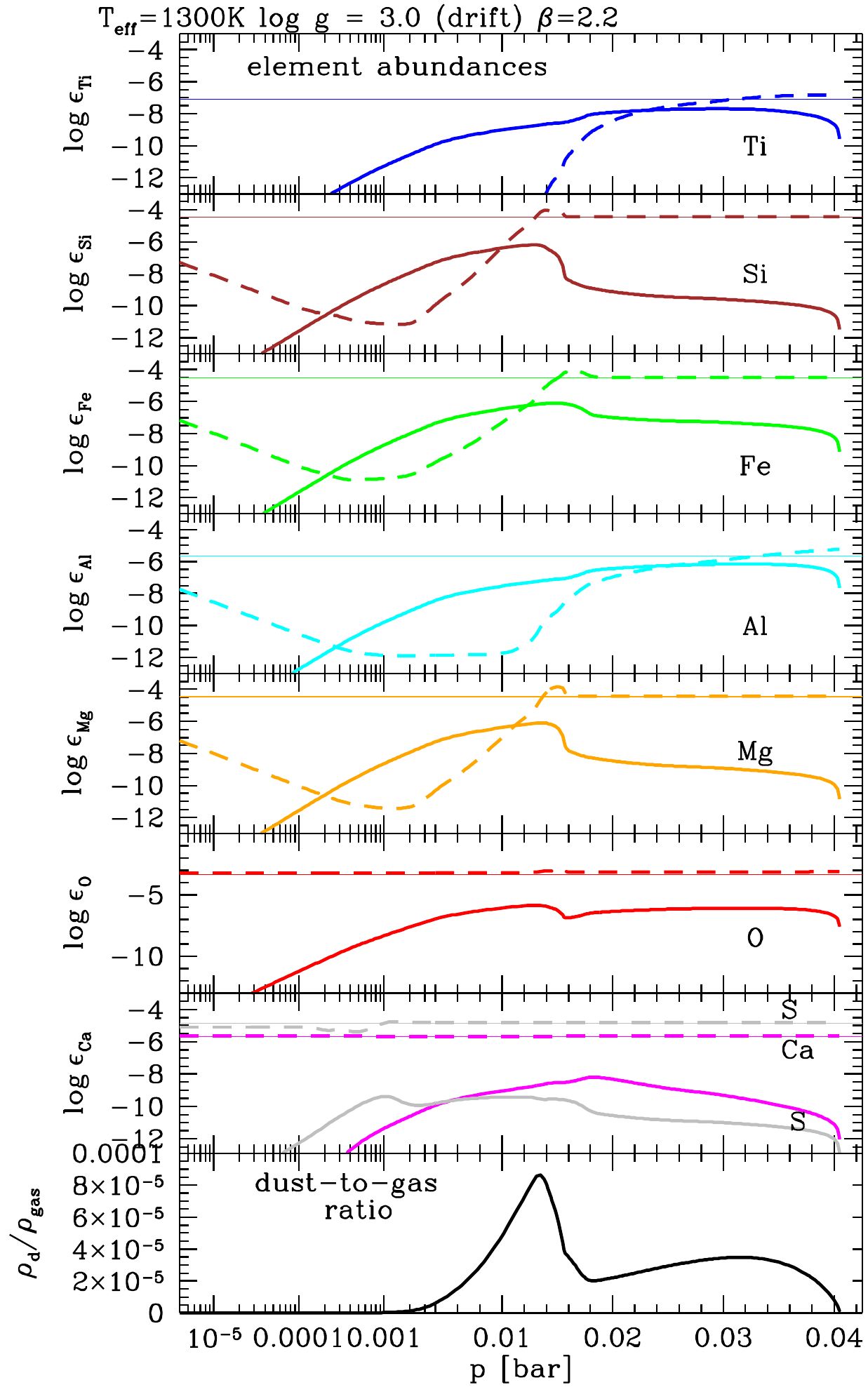}
\caption{Element abundances changing through mineral cloud formation
  in the atmosphere of a planet for the same model like in Fig.~\ref{fig:13003.0ref}. \newline
  Plotted are the initial solar abundances (thin solid line), the
  actual gas-phase element abundances (dashed line), and the element
  abundances locked into the dust (thick solid line). The dust-to-gas
  ratio, $\rho_{\rm d}/\rho_{\rm g}$, is depicted for comparison
  (lowest panel).}
\label{fig:eps13003.0ref}
\end{figure}

\paragraph{\it Approach used:} 
All following results were obtained by solving a set of
moment and element conservation equations (see \cite{wh03, hwt08,
  hw06}) for a prescribed model atmosphere structure.  The moment
equations describe the seed formation, growth/evaporation and
gravitational settling for mixed grains made of 12 material species
that form by 60 surface reactions. We use a {\sc Drift-Phoenix} model
structure for gas temperature, gas pressure and convective velocity as
input \cite{wit09}. {\sc Drift-Phoenix} models use the same set of cloud model equations, only the number of condensing materials in lower than considered in this paper. The initial values for the element abundances were
solar \cite{asp09} but do change due to cloud formation as shown in
Fig.~\ref{fig:eps13003.0ref}. We will depart from this assumption in
Sect.~\ref{s:cloudeps}.  We note that the {\sc Drift-Phoenix} model
structure are calculated without the effect of an external radiation source.

\subsection{Cloud formation processes}

Cloud formation is an intrinsic non-equilibrium process during which
the gas-phase constituents participate in a phase-transition from
which the cloud particles emerge. No cloud particle can {\it form} in
phase-equilibrium as the very nature of an equilibrium is to be the
minimum energy state of a system where the system feels very
comfortable in.

Cloud formation in extrasolar atmospheres necessarily starts with the
{\it formation of seed particles} because we cannot assume that the
planetary object has a crust from which sand or ash particles are
diffused upwards or injected into the atmosphere by volcanic
eruptions.  In terrestrial atmospheric literature, {\it seed
  particles} are referred to as {\it cloud condensation nuclei}. These
are particles able to promote droplet formation at terrestrial
atmospheric water supersaturation levels. In more general terms, such
seeds provide a surface onto which other material can condense more
easily as surface reactions are considerably more efficient than the
sum of chemical reactions leading to the formation of the
seed. Figure~\ref{fig:13003.0ref} summarizes the results of our cloud
model for one example atmosphere model. 

\paragraph{\it Formation of seed particles:}  
The formation of the first surface out of the gas phase proceeds by a
number of subsequent chemical reactions that eventually result in small
seed particles. Such a chain of chemical reactions can proceed by
adding the same molecular unit (=monomer) during each reaction step
(e.g. \cite{je00}) which is referred to as {\it homogeneous
  nucleation}. {\it Heterogeneous nucleation} occurs if different
monomer units participate in different reaction steps to form larger
molecules and eventually clusters (e.g. \cite{bro12, pl13}).  We apply
the concept of homogeneous nucleation to the formation of TiO$_2$ seed
particles.  Figure~\ref{fig:13003.0ref} (2$^{\rm nd}$ panel, left) shows that
the nucleation rate (J$_*$) peaks rather high in the atmosphere and
falls off towards higher gas temperature (1$^{\rm st}$ panel, left). The peak
of the cloud particle number density (n$_{\rm d}$) coincides with the
peak of the nucleation rate but the number density remains high
towards higher temperature. This is a clear sign that cloud particles
fall into the atmosphere and therefore do exist below the seed
formation region.

\paragraph{\it Growth, evaporation:}  
Growth and evaporation are surface reaction onto a surface or off a
surface, respectively. They are determined by the composition of the
gas phase that provides the number density for surface reactions and
leads to the formation of a substantial mantle of a grain or droplet
on top of the seed. This mantle determines the mass, volume and main
chemical composition of the cloud particles. Many materials can be
simultaneously thermally stable in a gas but these materials change
depending on the carbon-to-oxygen ratio and the abundance ratios of
other elements. In principle, all stable and supersaturated material
can grow simultaneously on a seed
particle. Figure~\ref{fig:13003.0ref} (5$^{\rm th}$ panel) shows the
effective supersaturation ratios (S$_{\rm eff}$, \cite{hwt08}) for all
materials involved here (TiO$_2$[s], SiO[s], SiO$_2$[s], Fe[s],
FeO[s], Fe$_2$O$_3$[s], FeS[s], MgO[s], MgSiO$_3$[s],
Mg$_2$SiO$_4$[s], Al$_2$O$_3$[s], CaTiO$_3$[s] with the corresponding
surface reactions as in Table 1 in \cite{hwt08}). This demonstrates
that all materials are supersaturated at high atmospheric layers and
subsequently achieve phase equilibrium deeper in the cloud: They grow until the gas-phase
has reached phase-equilibrium for these particular
materials. High-temperature condensates like Al$_2$O$_3$[s] (light
blue), TiO$_2$[s] (dark blue) and CaTiO$_3$[s] (magenta) reach S=1 at
considerably higher temperatures where they under-saturated eventually and
 evaporate. This is particularly interesting if solid
particles form from the gas-phase as the composing materials comprise
silicates, oxides, and iron compounds leading to the formation of
grains of mixed materials \cite{hefo13, hwt08, wit09, hr9}. The
material composition is shown in 4$^{\rm th}$ panel in
Fig.~\ref{fig:13003.0ref} which only depicts silicates in orange/brown
colors, Fe[s] and Fe-compounds in green, Al$_2$O$_3$[s] in light blue,
TiO$_2$[s] in dark blue, and CaTiO$_3$[s] in magenta.

\paragraph{\it Gravitational settling (rain-out):} 
The equilibrium between friction and gravity determine how fast
the cloud particles fall through the atmosphere (\cite{wh03}). The
cloud particles will continue to grow and to change their material
composition during their way into denser and warmer atmospheric layers
(Fig.~\ref{fig:13003.0ref}, 4$^{\rm th}$ panel). Panel 2 of the same figure
shows that the cloud expands well blow the region of efficient
nucleation, hence, these particle fall in from above. The mean
grain size (6$^{\rm th}$ panel) is determined by the different dust formation
processes that govern the dust formation at different sites in the
atmosphere: The upper part of the cloud is governed by the nucleation
process. Growth is very inefficient due to the low density of the
ambient gas, hence, the cloud particles remain very small
and haze-like. Once the particles sink into deeper layers, the
surface growth process dominates, resulting in  strongly increasing
grain sizes, until $p_{\rm gas}\!\sim\!0.001$\,bar in the model
depicted. It follows a small minimum where all the silicates
evaporate (compare 4$^{\rm th}$ panel). Fe[s]-growth picks up but the grain
size does not change considerably during their remaining path
through the atmosphere. A comparison with the drift velocity (6$^{\rm th}$
panel, right) shows that the grains fall with almost constant speed
through the atmosphere until they evaporate at the bottom of the
cloud. 

\paragraph{\it Element depletion due to cloud formation:} 
We have summarized the cloud formation processes above. The cloud
formation has a strong impact on the local chemistry by depleting
those elements that participate in the formation of the cloud
particles.  Figure~\ref{fig:eps13003.0ref}  demonstrates how the gas-phase element abundances
change compared to the initial, canonical solar value for all elements
that participate in the cloud particle formation in our model. The
comparison with the dust-to-gas ratio ($\rho_{\rm d}/\rho_{\rm g}$,
lowest panel) shows that most of the dust is present when most of the
elements Mg, Si and O are locked in dust. Figure~\ref{fig:eps13003.0ref}  also emphasises that a C/O ratio alone is not sufficient to characterise the element abundances of an objects as elements change their abundances individually according to their involvement in cloud formation (or ice condensation in the protoplanetary disk to start with).


\begin{figure}[htbp]
\centering
 \includegraphics[scale=0.65]{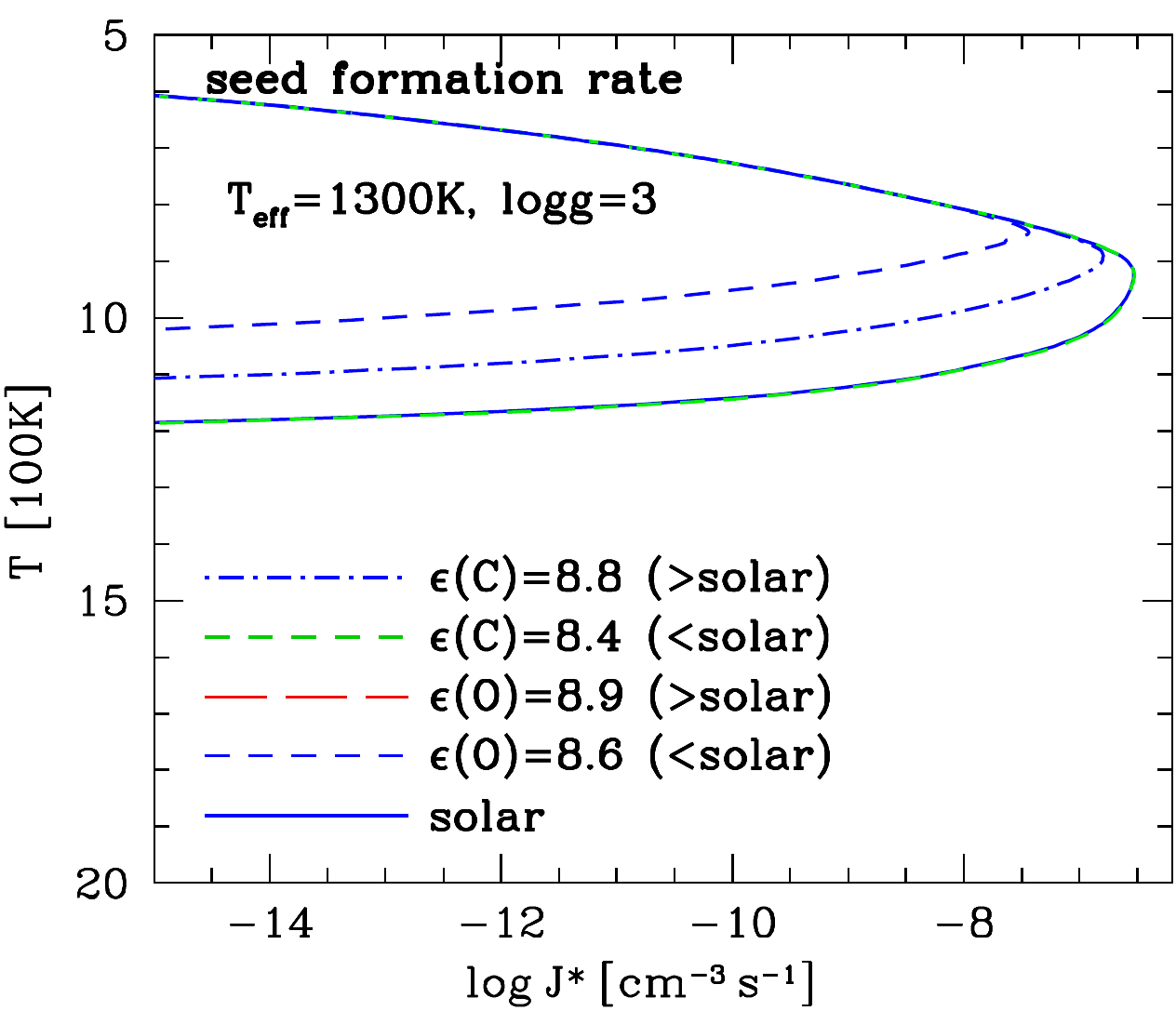}\\*[-0.2cm] 
 \includegraphics[scale=0.65]{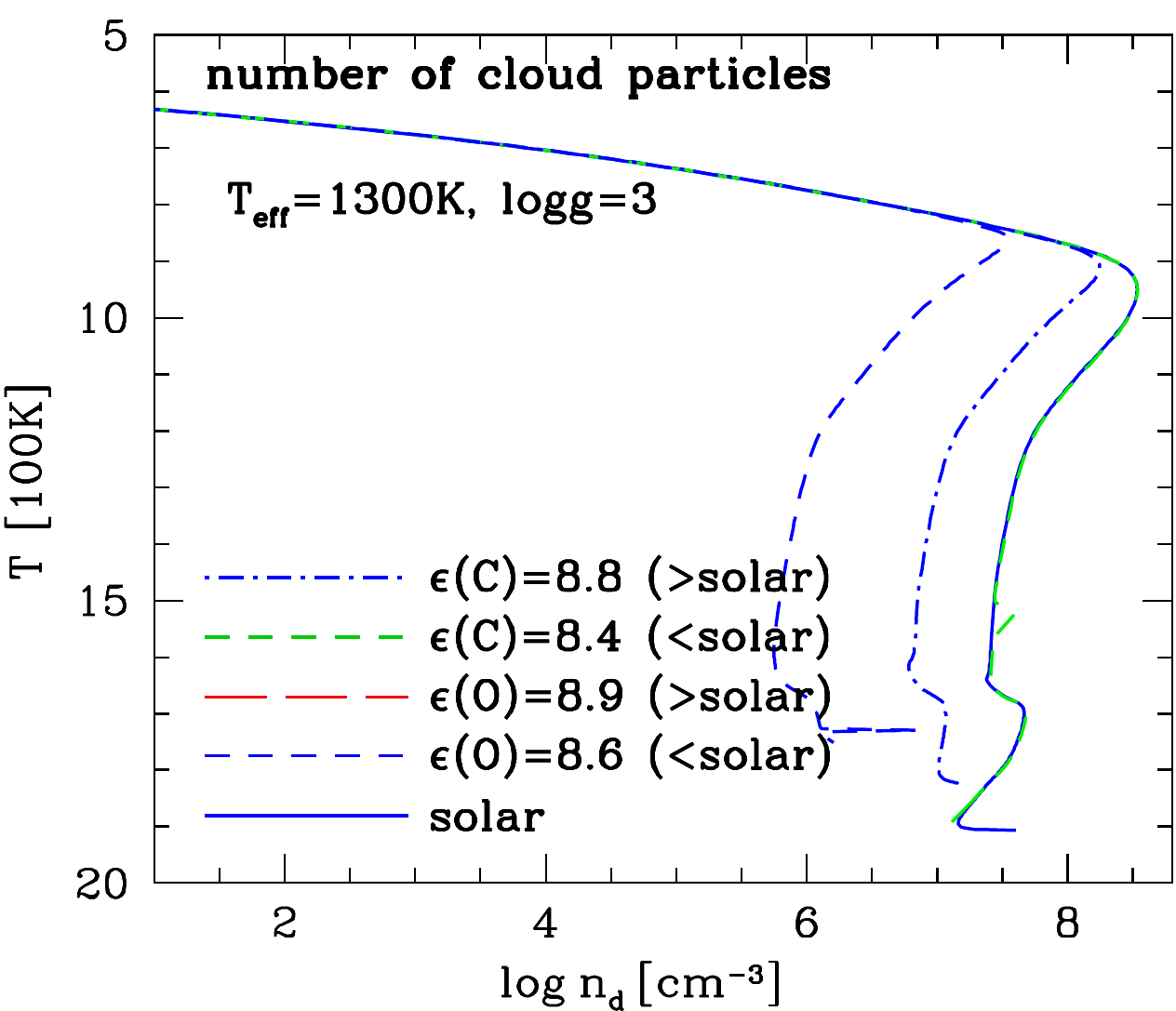}\\*[-0.2cm]
 \includegraphics[scale=0.65]{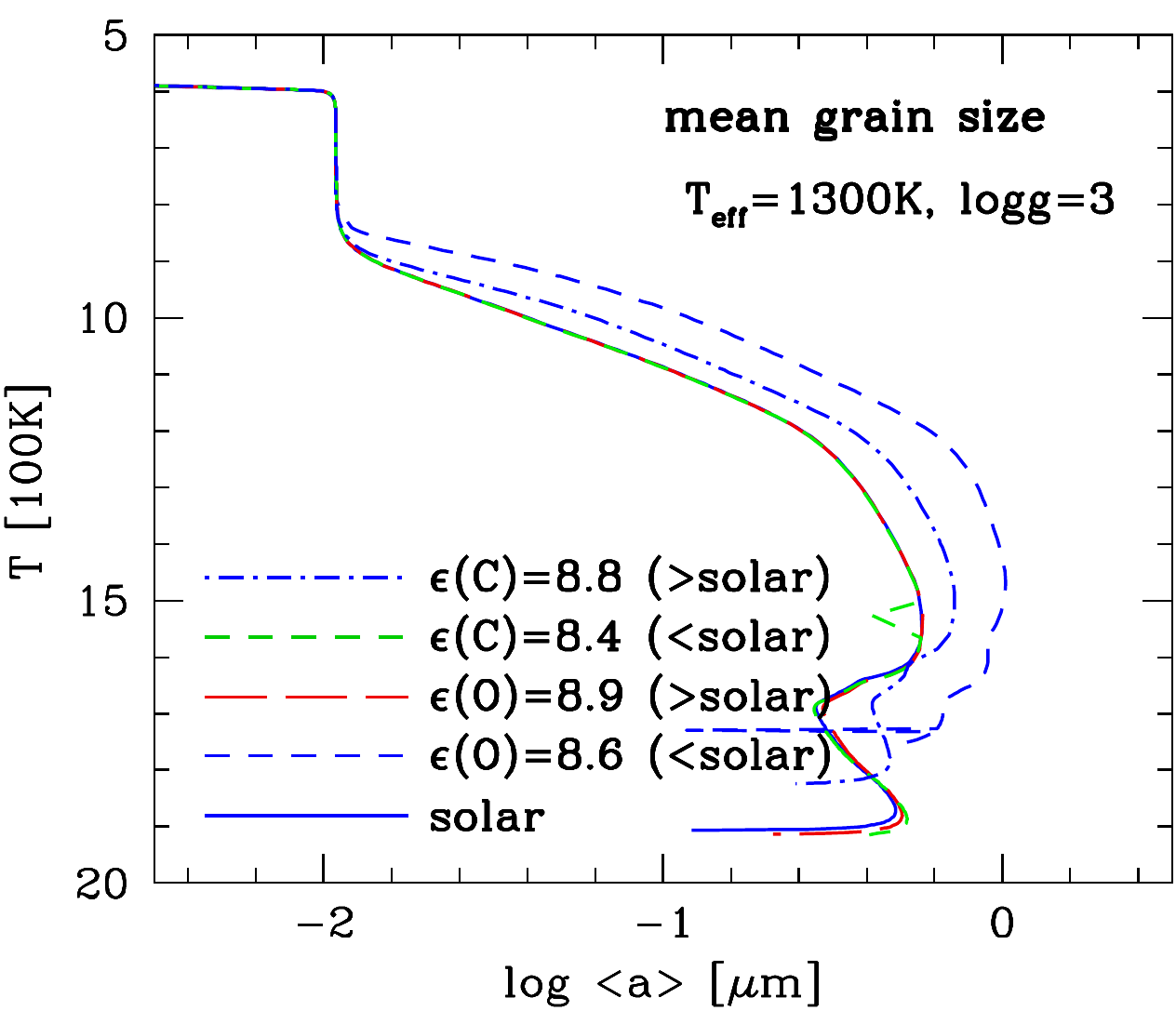}
\caption{Cloud properties for different oxygen ($\epsilon$(O)) or
  carbon ($\epsilon$(C)) abundances for a pre-scribed {\sc
    Drift-Phoenix} planetary model atmospheres (T$_{\rm eff}$=1300K,
  log(g)=3.0). {\bf Top:} nucleation rate, J$_*$ [cm$^{-3}$ s$^{-1}$];
  {\bf Middle:} number of cloud particles, n$_{\rm d}$ [cm$^{-3}$];
  {\bf Bottom:} mean size of cloud particles, $<$a$>$ [$\mu$m]. }
\label{fig:epsAtmos}
\end{figure}

\section{Changing cloud properties in atmospheres with non-solar abundances}\label{s:cloudeps}

 Cloud formation is determined by the local atmospheric properties,
 T$_{\rm gas}$ and $\rho_{\rm gas}$, and the local element abundances,
 $\epsilon_{\rm i}$. Cloud formation has a strong feedback on
 $\epsilon_{\rm i}$ due to element-dependent lock-up in the cloud
 particles, and on T$_{\rm gas}$ due to the dust opacity and due to
 element depletion that changes the gas opacity.  Studying cloud
 formation in atmospheres with non-solar abundances affords us a first
 assessment of how different the local chemistry might be in newly
 forming planets that are exposed to changing element abundances with
 distance from the star and during the history of disk evolution
 itself. It is of importance to understand to which extent element
 abundances influence the results of our cloud modeling and/or the
 local gas-phase chemistry, and how this might require a
 re-interpretation of, for example, the tentative detection of carbon
 planets or the conclusions draw from accumulating uncertainties in
 element depletion by phase-equilibrium cloud models, initial element
 abundances, mixing ratios and quenching heights \cite{bi2013,mos13,
   hell08b}.  We concentrate on the effect of element abundances only
 and do not consider irradiation, photo- or ion-chemistry, nor
 atmospheric circulation. Our results will therefore be directly
 applicable to planets in the outer portion of the disk where the host
 star's radiation field does not play a significant role. Here, the
 chemical composition in upper atmosphere above the cloud layer will
 be influences by cosmic ray chemistry \cite{rim14}.  Atmosphere
 models that take into account the presence of clouds in irradiated
 planets \cite{for08} consider a host star distance of $<0.05$AU which
 is in the oxygen-rich part of our disk models where oxygen-reducing
 ice formation does not play a role
 (Figs.~\ref{fig:epsDisk_UMIST2012},~\ref{fig:epsDisk_OSU2010}). The
 following study is applicable to directly imaged planets \cite{opp13,
   laf10, lec09, bon10, bon13, neu05, jas13} and free-floating
 planets, e.g.  \cite{del12, liu13}. The issue of non-solar element
 abundances is also relevant for the large number of close-in,
 irradiated planets as the element abundances are an essential input
 {\it and} output quantity.

\begin{figure}[htbp]
\centering
  \includegraphics[scale=0.84]{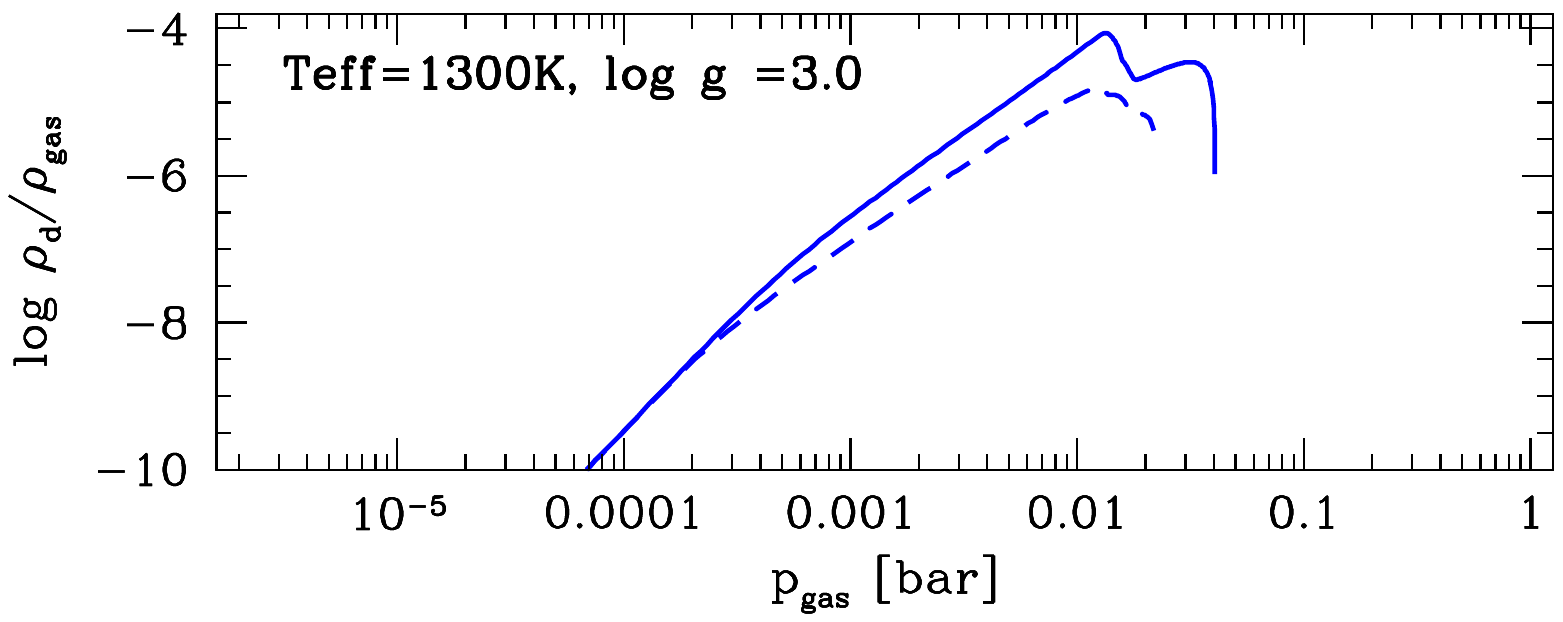}
\caption{Changing atmospheric dust-to-gas ratio with changing initial element
  abundances for models depicted in Fig.~\ref{fig:epsAtmos}. Shown 
  are the solar case (solid line) and the sub-solar case for
  $\epsilon_{\rm O}$=8.6 (dashed line). No differences in the $\rho_{\rm
    d}/\rho_{\rm g}$ ratio were found for the other cases.}
\label{fig:rdrgDisk}
\end{figure}

\subsection{Sub- and super-solar oxygen and carbon abundances}

After presenting the canonical result of our cloud formation model for
a planetary atmosphere with the solar element abundances as initial
values (Sect.~\ref{s:cloudmod}), we manipulate the initial values for
the oxygen and carbon abundances guided by the {\sc ProDiMo} disk
model results (Sect.~\ref{s:disk}). Our aim is to present a first
study of how much the cloud formation processes in a planetary
atmosphere would change for different C/O ratios as to be expected in
a protoplanetary disk.  As pointed out in Sect.~\ref{ss:results}, all
other elements can change too, but oxygen and carbon will dominate the
remaining disk chemistry.  Figures~\ref{fig:epsDisk_UMIST2012} and
\ref{fig:epsDisk_OSU2010} suggest that the C/O ratio changes with
disk age and with increasing radial distance from the star.  The
reason is the formation of water ice (H$_2$O\#) and CO-ice (CO\#) at
different radial distances  $>0.5$ AU ({\it ice lines}).  An almost pure H$_2$/He
gas is left already in young disks at radial distances beyond the
CO$_2$-ice line, hence, young planets forming at such distances would
have an extremely metal-poor atmosphere in contrast to their host
star's element composition.  At smaller radial distances from the star
$< 13$AU, the oxygen and carbon abundances change time-dependently in
the sandwich zone between the H$_2$O-ice and the CO$_2$-ice lines.

In order to understand which effects changing C/O ratios have on the cloud
structure, we consider lower ($\log\epsilon_{\rm O}$=8.6,
$\log\epsilon_{\rm C}$=8.4) and higher ($\log\epsilon_{\rm O}$=8.9,
$\log\epsilon_{\rm C}$=8.8) oxygen and carbon abundances compared to
the solar values ($\log\epsilon_{\rm O}^{\rm solar}$=8.87,
$\log\epsilon_{\rm C}^{\rm solar}$=8.55) that are commonly used in
model atmosphere simulations. All other input quantities are kept the
same. Element abundances are given relative to the hydrogen abundance
($\log\epsilon_{\rm H}$=12).

Figure~\ref{fig:epsAtmos} shows that fundamental cloud
properties change due only to  a change in oxygen/carbon abundance:
The seed formation rate (top panel, TiO$_2$ seeds) decreases with
decreasing oxygen abundance and increasing carbon abundance, and this has
fundamental implications for the cloud opacity as it changes the mean
grain size (bottom panel). Decreasing oxygen and increasing
carbon abundance have the same net effect of decreasing the amount of
oxygen available for TiO$_2$ molecules (seed monomer) to be formed. If more carbon is
available, more oxygen will be locked in carbon-monoxide (CO). Hence,
already small changes in $\epsilon_{\rm O}$ and $\epsilon_{\rm C}$
affect the Ti-chemistry because Ti has a very low element abundance (see
\cite{hw06} Fig.5).

The lower rate of seed formation results in less cloud particles being
formed (middle panel, Fig.~\ref{fig:epsAtmos}). Clouds in a low-oxygen
abundance gas (but still C/O$<$1; dashed and long-dashed line) do have
fewer cloud particles,  suggesting a more transparent haze layer.
These particles will rain into the denser atmosphere and grow to
bigger sizes than in the solar abundance case (solid line). The
maximum grain size increases from $0.4\mu$m in the solar case to about
$1\mu$m  due solely to a moderate decrease of the oxygen abundance (or
increase of $\epsilon_{\rm C}$). We note that an increase of
$\epsilon_{\rm O}$ to values larger than the solar values and a
decrease of $\epsilon_{\rm C}$ below the solar value does not have any
effect on the cloud properties. The reason is that all possible
binding partners (Si, Fe, Mg, Al) to O have much lower element
abundances and are already locked up in molecules. The CO molecule acts
as an oxygen-sink and no increase of carbon could change this under the
conditions of gas-phase chemical equilibrium applied here. This leads
to the conclusion that the cloud properties would change further if
the Si, Fe, Mg, Al abundances would change too. At a first glance,
this would mean that less material could grow onto seed
particles. However, the gas-phase chemistry may offer other surface
reactions than those used in our cloud formation model (Table 1,
\cite{hwt08}).

We test the implication of the changing effective oxygen abundance on
the dust-to-gas ratio, $\rho_{\rm d}/\rho_{\rm g}$, in planetary
atmospheres (Figure~\ref{fig:rdrgDisk}).  The largest change of about
a factor of 10 occurs for a decreasing oxygen abundance which supports
our conclusions from Figs.~\ref{fig:epsDisk_UMIST2012} and
\ref{fig:epsDisk_OSU2010} above.

\subsection{Sub-solar oxygen and carbon abundances at the extreme: C/O=0.99}

\begin{figure}[htbp]
\centering 
\includegraphics[scale=0.6]{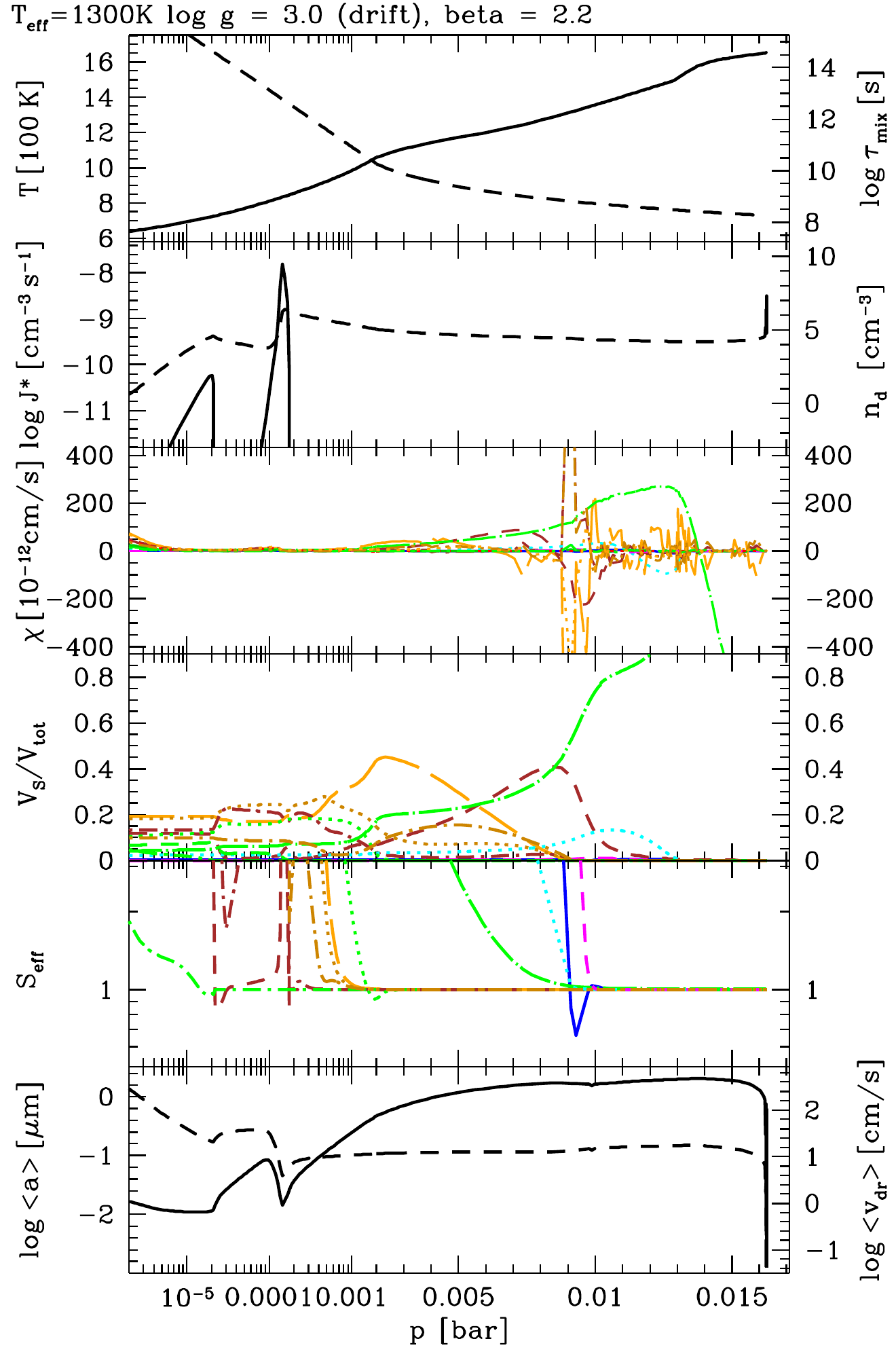}
\includegraphics[scale=0.6]{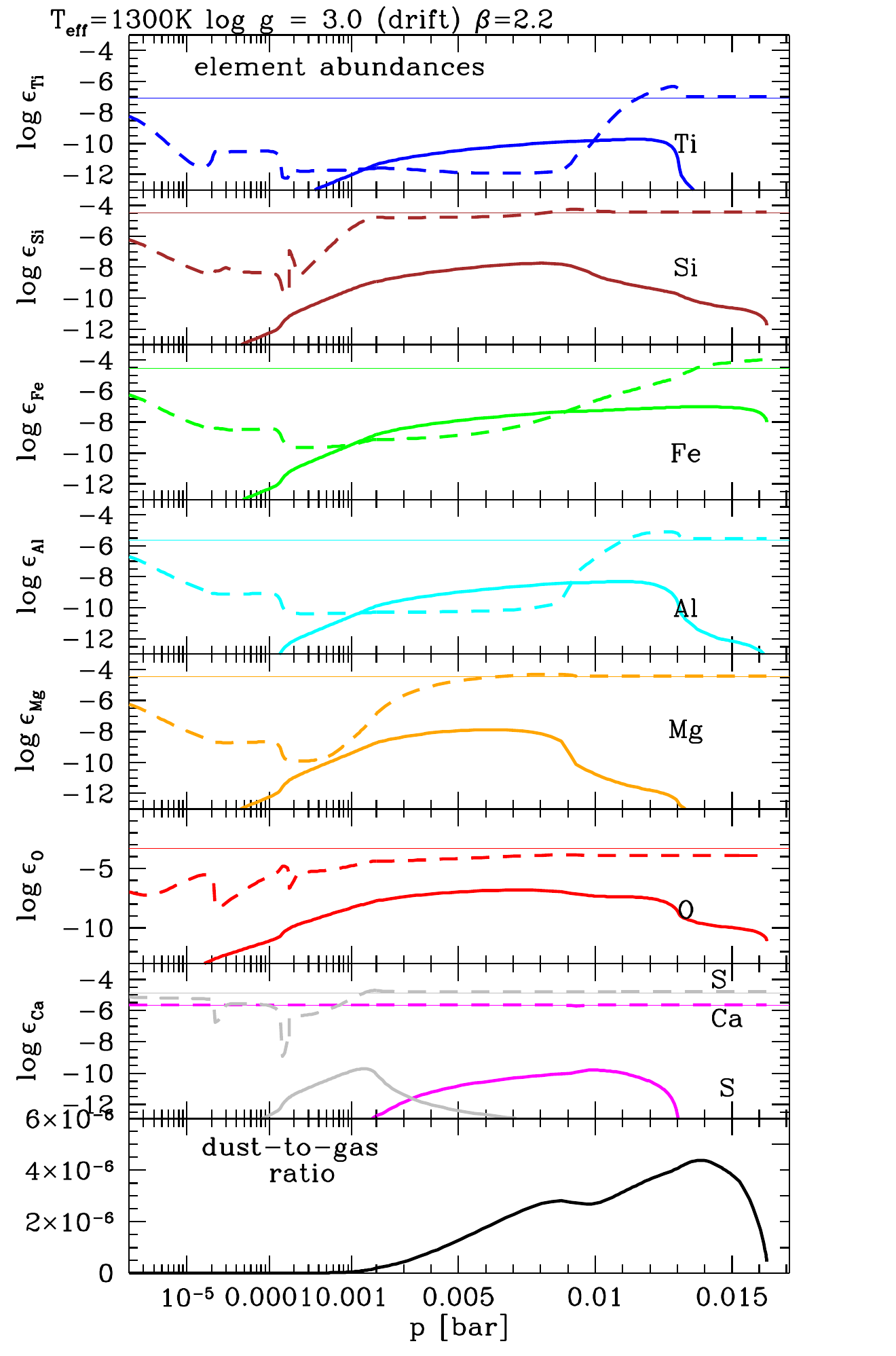}
\caption{Cloud structure results for a gas of C/O$_{\rm init}$=0.99 ($\epsilon_{\rm
    O}=8.07$, $\epsilon_{\rm C}=8.06$) for the same {\sc
    Drift-Phoenix} atmosphere model as in the previous
  figures. {\bf Left:} Cloud structure and physical properties.  {\bf Right:} Element abundances changing through mineral
  cloud formation. In both figures, the same line coding is used as in Figs.~\ref{fig:13003.0ref},~\ref{fig:eps13003.0ref}.}
\label{fig:130030_disk2}
\end{figure}

The outstanding question is if carbon-rich planets could exist around
oxygen-rich host stars. The core-accretion scenario for planet
formation does not presently suggest the in-situ formation of
carbon-rich planets, unless models are very much simplified to match
the observations. We have shown that cloud formation might play a
crucial role in tipping over an oxygen-rich atmospheric gas
composition locally towards a carbon-rich atmospheric gas,  due to
  the additional formation of oxygen-rich dust in the atmosphere.  We
have tested this hypothesis by decreasing the oxygen abundances in our
cloud formation {\sc Drift} code to values as low as suggested by {\sc
  ProDiMo} ($\epsilon_{\rm O}\!=\!8.07$, $\epsilon_{\rm C}\!=\!8.06$)
approaching $\rm C/O\!\approx\!1$. Figure~\ref{fig:130030_disk2} shows
the results for the cloud structure (left) and the element abundances
(right): The seed formation rate is very low compared to all results
from the previous sections due to the much lower oxygen
abundance. TiO$_2$-seed formation is still possible due to a small
surplus of oxygen that is not locked by CO. But considerably less
particles are formed and two detached nucleation maxima determine the
number of cloud particles. The upper most nucleation event is killed
off by the oxygen-consumption during the growth process involving
TiO-molecules. Nucleation resumes only where the atmospheric density
is sufficiently high and the temperature sufficiently low. Note that
Fig.~\ref{fig:130030_disk2} (left) shows the peak values of the
nucleation rate, and that the atmospheric range affected by nucleation
and growth extends towards considerably lower pressures (compare
Fig.~\ref{fig:130030_disk2}, right).
The competition for condensable material become apparent from the
double-peaked nucleation rate $J_*$: The growth of silicates consumed
oxygen which decreases the number of TiO$_2$ molecules in the gas,
causing the nucleation process to stop. A local increase of the mean
grain size results and the grains form a semi-detached haze layer of
0.1 $\mu$m silicate grains. The second nucleation peak coincides with
a decreasing grain size just on top of a deeper cloud layer of an
almost constant grain size of 1$\mu$m.

\begin{figure}[htbp]
\centering 
\includegraphics[scale=0.9]{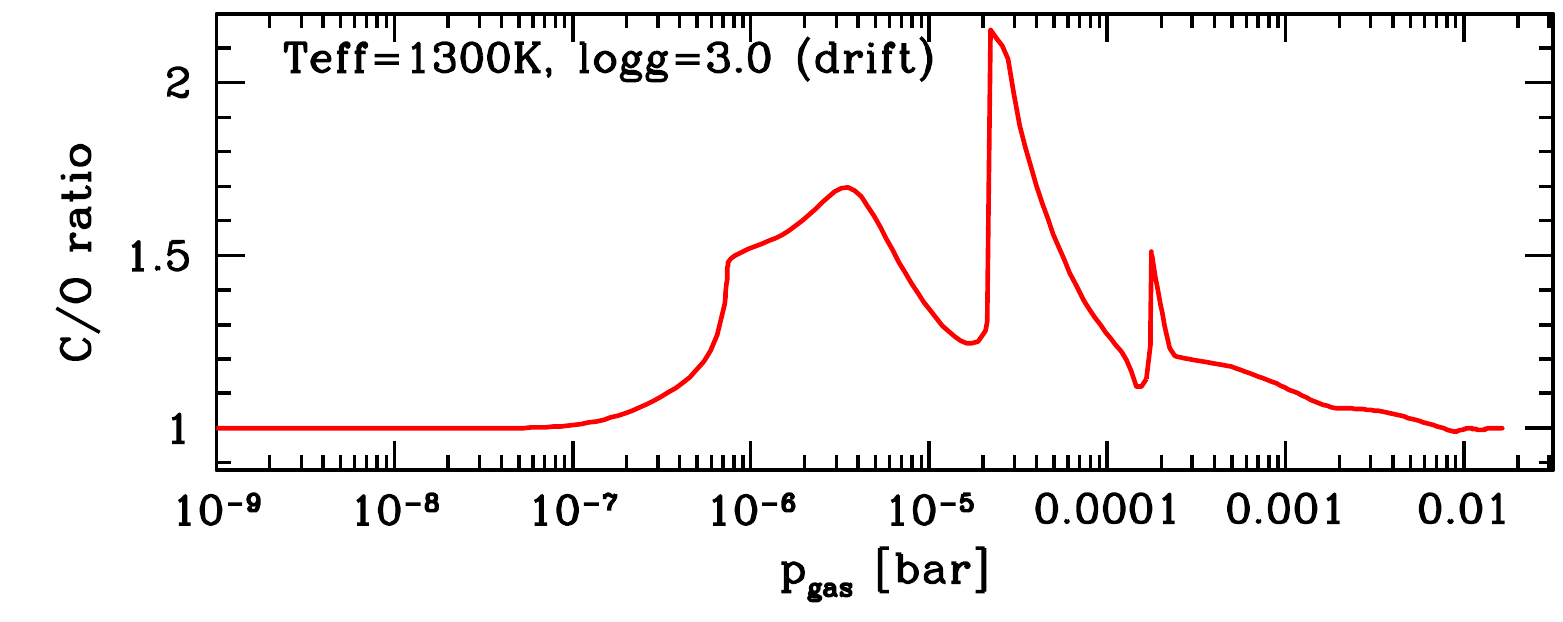}
\caption{C/O ratio after cloud formation from an initial element
  abundance with C/O$_{\rm init}$=0.99 ($\epsilon_{\rm O}=8.07$, $\epsilon_{\rm
    C}=8.06$; all other elements solar). The oxygen depletion by cloud
  formation clearly tips the gas-phase from an oxygen-dominated to a
  carbon-dominated chemistry.}
\label{fig:130030_CzuO}
\end{figure}

\subsection{Implications for the chemical composition of the atmosphere}

The gas-phase composition is determined by the local temperature (and
density) and the local element abundances. Both are affected by cloud
formation since cloud formation causes a considerable element
depletion of the gas phase. The C/O-ratio of a cloud forming
atmosphere with an initial C/O$_{\rm init}=0.99$ can increase to
values C/O$>1\,\ldots\,2$ alone due to oxygen depletion by cloud
formation (Fig.~\ref{fig:130030_CzuO}). As a result of cloud
  formation, the atmosphere can become locally very carbon-rich. We, however, do not find  C/O ratios $\sim$100. Our
  results suggest further that young planets that accrete their first
  atmosphere from the disk dust and gas will most likely have a
  strongly oxygen-depleted. This is because the primordial gas will
  condense more easily onto the previous disk grains due to an
  increased local density during accretion.  Cosmic Rays can enhance
  the fraction of hydro-carbon molecules (via ion-neutral reaction)
  \cite{rim14}, a process that can be considerably more efficient in
  an atmosphere that is oxygen-depleted as result of disk evolution
  and cloud formation.

\begin{figure}[htbp]
\centering 
{\ }\\*[-2.5cm]
\hspace*{-1cm}\includegraphics[scale=0.55]{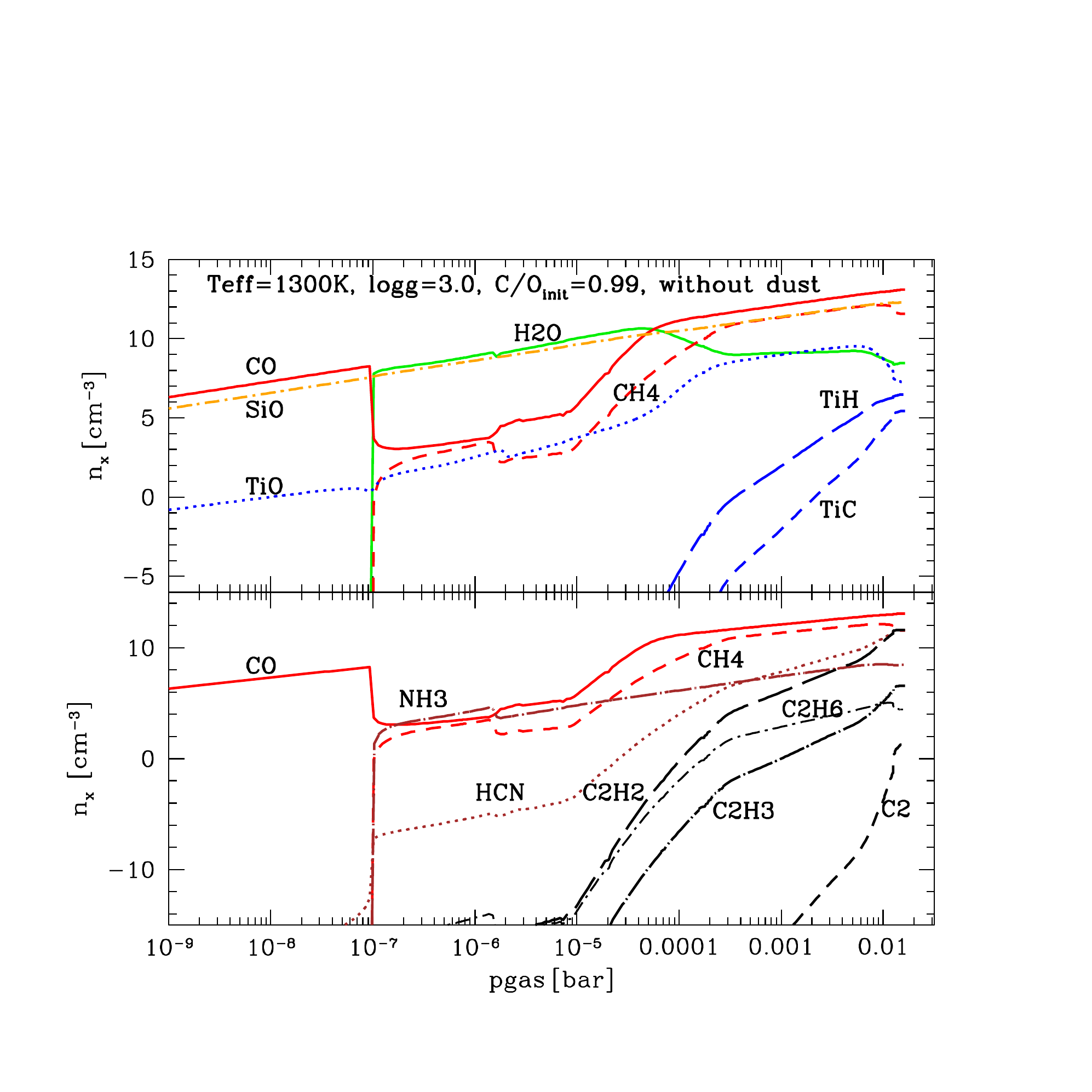}
\hspace*{-2cm}\includegraphics[scale=0.55]{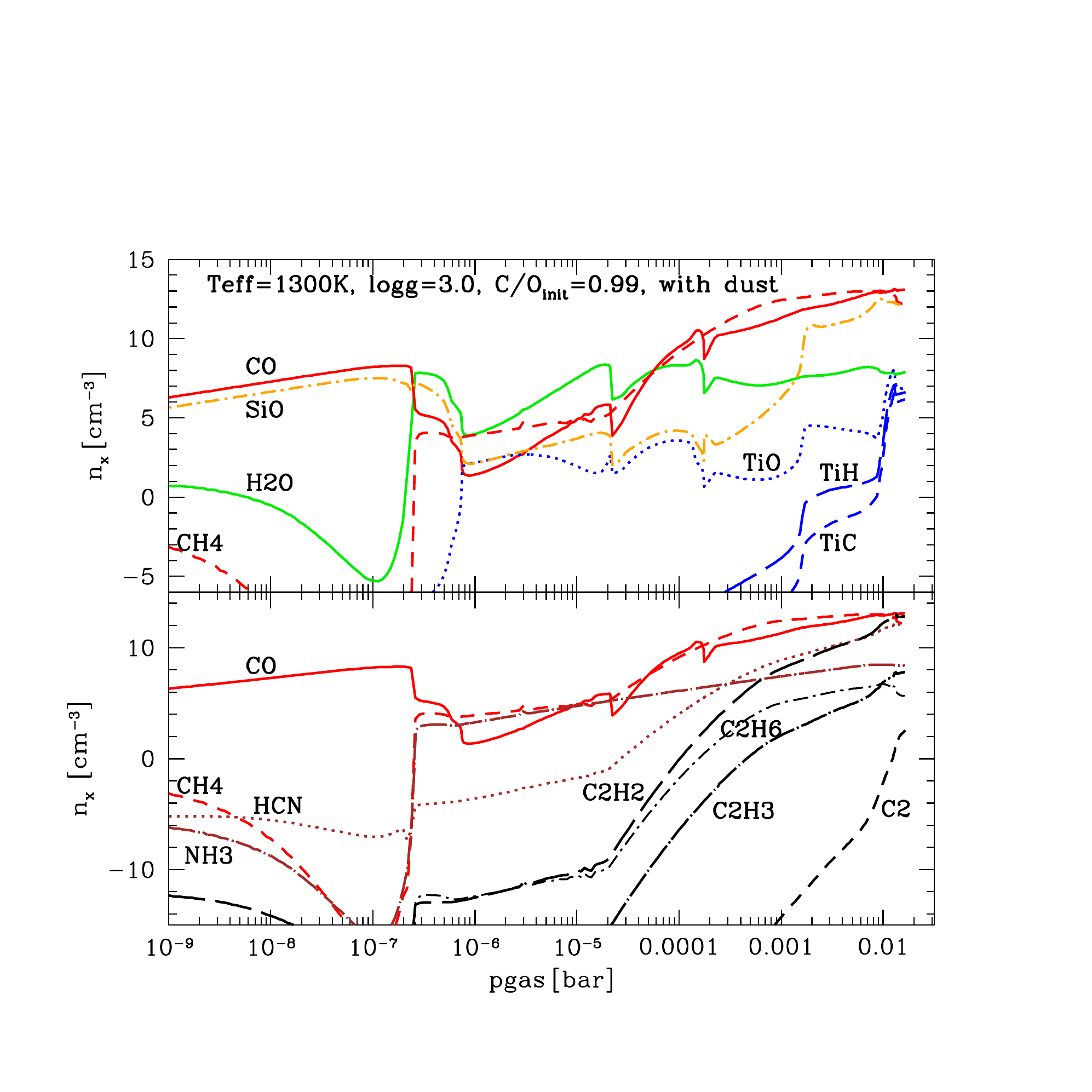}
\caption{Atmospheric molecular number densities in chemical equilibrium
  for the planetary model atmosphere in
  Fig.~\ref{fig:130030_disk2}. {\bf Left:} No element depletion by
  dust formation (C/O$_{\rm init}=0.99$), {\bf Right:} With element depletion by dust
  formation (O, Si, Mg, Fe, Ti, Ca, Al) resulting in the C/O ratio
  depicted in Fig.~\ref{fig:130030_CzuO} (C/O$>1\,\ldots\,2$). }
\label{fig:130030_gg}
\end{figure}

We have performed a calculation of the gas-phase composition of the
atmosphere depicted in Fig.~\ref{fig:130030_disk2} by applying our
chemical equilibrium code \cite{bi2013}. This code allows
us to see how the abundance of molecules that are strong opacity
carriers in oxygen-rich atmosphere (H$_2$O, CO, SiO, TiO) change if
the local C/O ratio changes. We also consider CH$_4$ and NH$_3$ which
are discussed as possible bio-marker molecules, and hydrocarbon chains and
cyano-molecules. The abundance of these molecules is shown in the left
panel of Fig.~\ref{fig:130030_gg} for C/O$_{\rm init}=0.99$ with no
element depletion by dust formation as reference values for the case
with element depletion by dust formation (right pane) causing
C/O$>1$. In the left panel, the dust only influences the
local temperature structure causing the drop of CO at $~10^{-7}$~bar
in the model results shown. We note the increasing abundance of
hydrocarbon chains (C$_2$H$_2$, C$_2$H$_6$, C$_2$H$_3$) and HCN with
increasing pressure below the cloud layer.  The molecular chemical
equilibrium abundances for a gas with C/O close to unity (but not quit
1) is very sensitive to small changes in the oxygen/carbon
abundance. This result is not new and was discussed for S-type AGB stars by
e.g. \cite{jo1993}. Hydrocarbon absorption were observed in S-type AGB
stars while it is surprisingly difficult to identify PAH absorption
feature in carbon-rich AGB stars, e.g. \cite{smolt2010}.

The number densities of all typical oxygen-rich gas phase molecules
decreases considerably with C/O$>1$. In the cloud layers, H$_2$O
remains the most abundant species after H$_2$. The H$_2$O, CO and the SiO gas
abundances in particular are a negative fingerprint of the dust growth
process (e.g. 4$^{\rm th}$ panel, left of Fig.~\ref{fig:130030_disk2}).  CH$_4$
and NH$_3$ become more abundant than CO in the cloud formation zone 
because it is not affected by the element depletion due to cloud
formation.  As before, NH$_3$ and CH$_4$ are of similar abundance, but
CH$_4$ is somewhat more abundant than NH$_3$. The largest difference
between the depleted (right of Fig. \ref{fig:130030_gg}) and the
un-depleted case (left of Fig. \ref{fig:130030_gg}) is the increasing
abundances of H$_2$O, CH$_4$, HCN, NH$_3$, C$_2$H$_2$ and C$_2$H$_6$
in the low-pressure regime near the cloud top. HCN becomes relatively
more important which is typical for low-metallicity gases near
C/O$=1$.

\section{Conclusions}

\noindent We have discussed the chemical preconditions for planet
formation in protoplanetary disks, with emphasis on the time-dependent
segregation of carbon, nitrogen and oxygen into gas and ice phases. 
We obtained the following results:
\begin{itemize}
\item The segregation into gas and ice phases beyond the water ice-line
  (the ``snowline'') result in a rich variety of gaseous oxygen,
  carbon, and nitrogen abundances in the midplanes of protoplanetary disks,
  depending on time, position in the disk, and cosmic ray ionization
  rate. The resulting gas element abundances can vastly differ from
  that of the host star.

\item Inside of the snowline ($\gtrsim\!150\,$K) all considered 
  ice phases are thermally unstable, and the gas phase abundances
  remain primordial. 

\item  Beyond the CO ice-line ($\lesssim\!20\,$K) oxygen, carbon
  and nitrogen freeze out quickly, and already after
  $\ll\!10^3\,$yrs, the outer midplane barely contains any molecules
  other than H$_2$. This may be different though, for the outermost 
  midplane which is transparent to interstellar UV and X-ray
  irradiation, as well as for scattered stellar UV and X-ray
  irradiation.

\item Between the snowline and the CO ice-line, a slow transition from
  O-rich to $\rm C/O\!\to\!1$ takes place, on timescales of
  $\sim$\,3\,Myrs.  This timescale is related to the cosmic-ray
  induced un-blocking of O$_2$ and CO, and scales with the cosmic ray
  ionization rate assumed.

\item For very long-lived protoplanetary disks, or disks exposed to
  an unusually high cosmic ray ionization rate, the carbon-to-oxygen
  ratio C/O would eventually exceed unity, leading to a sudden
  occurrence of organic molecules in the midplane, and providing the
  chemical pre-conditions for the formation of carbon planets.

\end{itemize}
Following the standard core-accretion model, it is this
element-depleted gas that will be finally accreted onto the
proto-planet in a rapid run-away phase, to eventually form the
planetary atmosphere, although many difficult questions remain open,
like the subsequent bombardment with left-over planetesimals, or the
previous internal heating and outgasing of the planetesimals due to
radioactive decay of $^{26}$Al.  However, in agreement with
  \cite{Oberg2011}, we conclude that a super-solar C/O ratio (but $\rm
  C/O\!\lesssim\!1$) is the most likely result from the gas-ice
  segregation in the disk, between the snowline and the CO ice-line.

In the remainder of this paper, we have studied how the cloud
formation in young planetary atmospheres is affected by decreased
oxygen abundances. Our results are applicable to directly imaged
planets like HB~8799b,c,d,e, GQ Lupi or $\beta$ Pic b, and
free-floating planets. But the issue of non-solar element abundances
has biased also our understanding of the atmospheres of the large
number of close-in, irradiated planets. Cloud formation depends
strongly on the local element abundances involved (Fe, Ti, Si, O,
$\ldots$) and it strongly affects the local elements by element
depletion or enrichment. The consequence is a strong influence of the
cloud formation on the local opacity, and hence on the atmosphere's
temperature structure. The oxygen abundance has a strong impact on the
seed particle formation rate which initiates the cloud formation
process. The reduced number of seed particles leads to a lower number
density of cloud particles which hence grow to larger sizes throughout
most of the cloud in an atmosphere with $\rm C/O\!\lesssim\!1$. This
sequence of processes leads to a more transparent haze layer on
low-C/O planet compared to a solar abundance (or solar-abdundance
scaled) atmospheres. However, an increasing oxygen abundance does not
automatically cause more cloud particles to be formed because the
nucleation rate is determined by the monomer density, not by the
oxygen abundance alone. We further observe the appearance of a
semi-detached cloud layer with C/O$\rightarrow$1.

We conclude that the differences in element abundances with radial
distance in protoplanetary disk have broad implications for the cloud
properties in planetary atmospheres. The element abundances are, however not a
  multiple of the set of solar abundances. Using the results of our disk
  models as input for our cloud and chemistry calculation did only
  result in C/O $\approx 2$, and and we can not confirm element abundances in planetary atmospheres as
  high as 100x the solar values. Planetary atmospheres might
  still carry signatures of the initial abundances of the gas in the
  protoplanetary disks from which they were once formed. However, it
  seems difficult to detect these signatures from spectroscopy
  directly, because cloud formation changes the gaseous element
  abundances.

We further have demonstrated that it is not straight-forward to argue for the
formation of carbon-rich planets (Figs.~\ref{fig:epsDisk_UMIST2012},~\ref{fig:epsDisk_OSU2010}) and
that some fine-tuning would be necessary in order to end up with a
planet that exhibits spectral signatures typical for a carbon-rich
gas.  So far, only additional processes like element depletion by dust
cloud formation and condensation adequately explain observations of
planetary atmospheres rich in carbon-binding molecules.


\acknowledgements{Acknowledgments} 
\noindent We thank the two referees for their constructive criticism which helped us improving the paper.
ChH and PBR highlights financial support
of the European Community under the FP7 program by an ERC starting
grant.  PW, IK, and WFT  acknowledge funding from the European Union Seventh
Framework Programme FP7-2011 under grant agreement no 284405.  Most
literature search was performed using the NASA Astrophysics Data
System ADS.  Our local computer support is highly acknowledged.



\bibliographystyle{mdpi}
\bibliography{bibCh}

\end{document}